\documentclass{article}

\usepackage{PRIMEarxiv}

\usepackage[utf8]{inputenc} % allow utf-8 input
\usepackage[T1]{fontenc}    % use 8-bit T1 fonts
\usepackage{hyperref}       % hyperlinks
\usepackage{url}            % simple URL typesetting
\usepackage{booktabs}       % professional-quality tables
\usepackage{amsfonts}       % blackboard math symbols
\usepackage{nicefrac}       % compact symbols for 1/2, etc.
\usepackage{microtype}      % microtypography
\usepackage{xcolor}         % color text
\usepackage{graphicx}
\graphicspath{{Image_Delay-Based-PUFs/}}     % organize your images and other figures under media/ folder
\usepackage{subfigure}
  \usepackage{algorithm}
\usepackage{algorithmic}
\usepackage{float}
\usepackage{tabularx}
\usepackage{multirow}
\usepackage{enumitem}
\usepackage{wrapfig}
\title{Exploring Delay-based PUFs for Energy-Efficient Low-Overhead Security of Wearable Devices}

\author{
  Venkata Prasanth Yanambaka \\
  School of Sciences \\
  Texas Woman's University\\
  Denton, TX, USA, 76201\\
  \texttt{vyanambaka@twu.edu} \\
  \AND
  Uma Choppali \\
  Department of Physics\\
  Dallas College - Eastfield Campus, \\
  Mesquite, TX, USA 75150 \\
  \texttt{uchoppali@dallascollege.edu} \\
   \And
  Saraju P. Mohanty \\
  Department of Computer Science\\
  University of North Texas \\
  Denton, TX, USA, 76201\\
  \texttt{smohanty@ieee.org} \\
}

\begin{document}
\maketitle

\begin{abstract}
The Internet of Things (IoT) was introduced almost two decades ago. In the past two decades, technology has seen huge advancements. Many devices have become powerful and have less power consumption. Many IoT architectures and environments were introduced to help make life easier, especially in wearable devices. The market for these wearable devices has constantly increased over the years and is expected to reach its maximum in the next couple of years. They also pose a threat to users' privacy and security because they constantly store and transmit personal information such as location, heart rate, and other sensitive data. Therefore, addressing the security vulnerabilities is a crucial aspect of this research. This paper presents a hardware-assisted, energy-efficient, low-overhead security solution for wearable devices. Specifically, two Physical Unclonable Function (PUF) architectures: Arbiter PUF and Hybrid Oscillator Arbiter (HOA) PUF are analyzed for integration in IoT systems. The result shows that Arbiter PUF consumes 25 $\mu$W, whereas HOA PUF consumes only 2.7 $\mu$W to generate keys for cryptographic purposes. These architectures introduce minimal power overhead while providing robust security, making them well suited for resource-constrained IoT ecosystems.
\end{abstract}

% keywords can be removed

\section{Introduction}
\label{SEC:intro}

Internet of Things (IoT) plays a vital role in shaping the modern world \cite{Mohanty_Book_2015_Mixed-Signal}. With rapid advancements in Artificial Intelligence (AI), the implementation of complex IoT architectures has become significantly more accessible and efficient. Robotics, in particular, integrates seamlessly with the IoT ecosystem, enabling the deployment of autonomous systems across smart healthcare, intelligent cities, and industrial automation. Autonomous drones, for example, are increasingly being deployed in diverse domains ranging from large-scale public events to military operations \cite{Schuck_2025_IEEE_Robotics}. Furthermore, Large Language Models (LLMs) accelerate IoT development by assisting with code generation, system design, and predictive analytics, making these technologies more accessible to developers across varying expertise levels. Recent market forecasts estimate that approximately 9.1 billion active autonomous "smart" devices (AIoT) will be active worldwide by 2033. This exponential growth is anticipated to accelerate the transition toward Industry 5.0, emphasizing human-centric manufacturing, sustainability, and resilience through the integration of advanced digital technologies \cite{EuropeanCommission2022Industry50}.

The rapid proliferation of AI-enabled IoT systems also expands the attack surface, thereby increasing system vulnerabilities. Recent studies have demonstrated that advanced AI agents can identify and exploit software weaknesses with increasing sophistication \cite{anthropic2026fable5statement}. Even with protective guardrails in place, many users actively attempt to jailbreak and test AI models to identify exploitable security gaps. To address these integration challenges, Anthropic developed an open-source protocol called the Model Context Protocol (MCP) designed to safely connect external tools, applications, and data sources to AI models. However, there have been numerous cases of AI agents being misused, leading to unauthorized access, security breaches, and cyberattacks \cite{khandelwal2026mcp_poisoning}.

% In the beginning of evolution, communication played a vital role. The evolution of communication since its origin has seen its major share of development in the past two decades \cite{Mohanty_Book_2015_Mixed-Signal}. It started around the time the term Internet of Things (IoT) was coined \cite{autoidlabs}. It has been a little over two decades since the term IoT was coined. In an IoT environment, ``Everything'' collects, ``Everything'' transmits and ``Everything'' is connected. With the research community focusing on IoT, many architectures tailored for specific applications were proposed over the years. Every device in the IoT network is required to have a unique identification \cite{Iqbal_IEEEIoT_2020}. IoT is considered as one of the six most ``Disruptive Civil Technologies'' by the National Intelligence Council, USA \cite{NIC_US_Impacts}.

% Many applications have been integrated with IoT architectures. Many areas such as Military \cite{Hoang_IEEEWC_2020}, Smart City \cite{Mohanty_CEM_2016-Jul}, Smart Farming \cite{Gupta_IEEEAccess_2020} and so on. Every device in these IoT architectures constantly transmit data over the network to the cloud or to an other node. Many off-the-shelf development boards have gained the communication capability. This gave the researchers to prototype any proposed architectures of IoT.

The rapid development in AI, and IoT devices has been made possible only with the shrinking transisor size on a chip \cite{Mohanty_Book_2015_Mixed-Signal}. Requirement for very high performance and ultra low power devices has never been so high. In recent years, AI has driven the demand for high-performance computing chips, including those equipped with high-bandwidth memory. The memory bandwidth for Nvidia H200 chips is 4.8 TB/s \cite{nvidia2024h200} which has risen from 336.5 GB/s on its Nvidia Titan X graphics processing units (GPUs) in 2015. Fin Field-Effect Transistors (FinFETs), 3D transistors that employ fin-shaped channel structures, has enabled to continue miniaturizing transistor dimensions to advanced technology nodes. Scaling bulk CMOS or planar transistors has been challenging beyond 32nm. But scaling resumed beyond 22nm with the FinFETs architecture and geometry. These advancements have enabled the development of high-performance, ultra-low-power devices for embedded and IoT applications.

\begin{figure}[hb]
	\centering
	\includegraphics[width=0.74\textwidth]{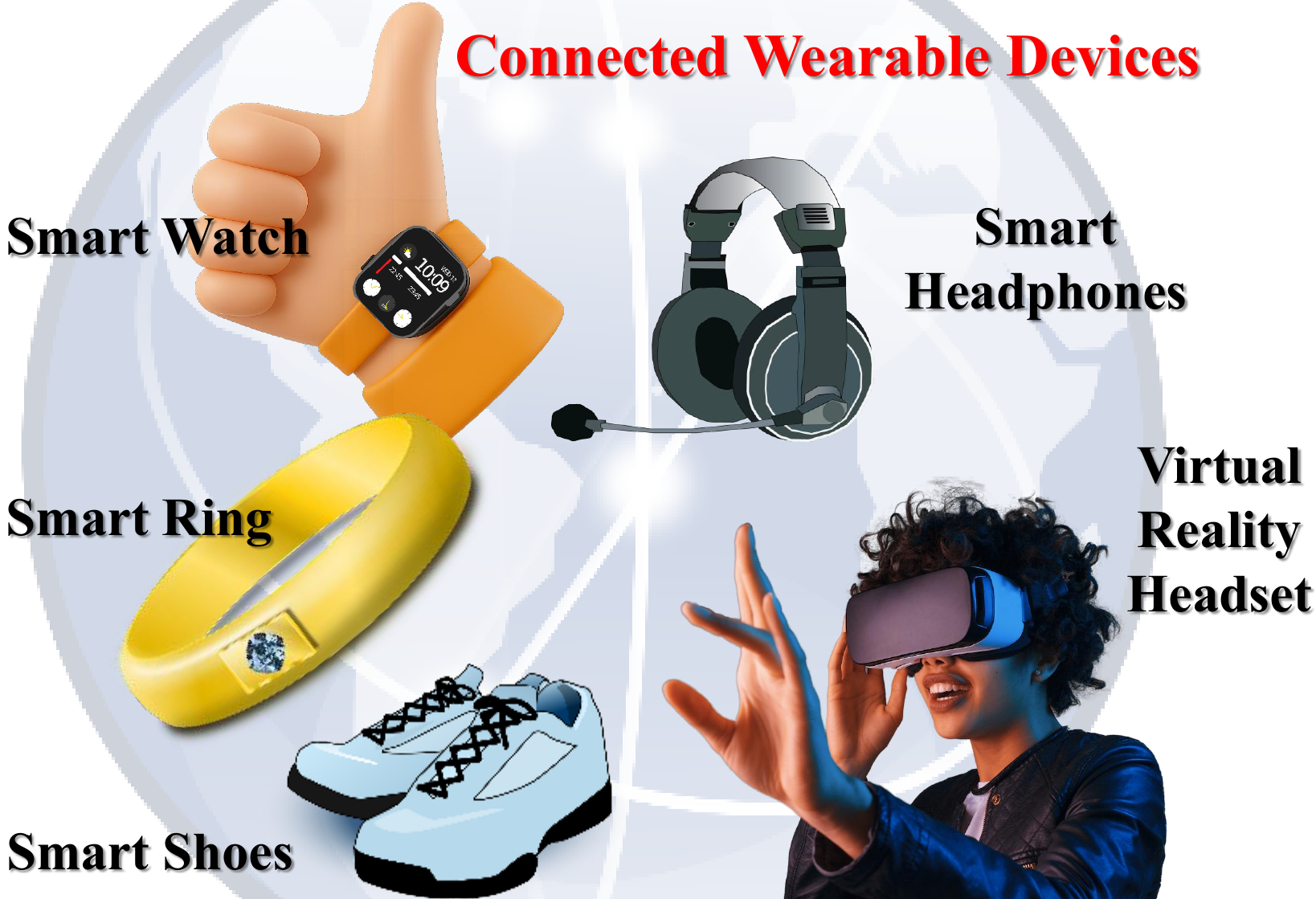}
	\caption{Illustration of the Internet of Things (IoT).}
	\label{FIG:IoT}
\end{figure}

Fig. \ref{FIG:IoT} illustrates the architecture of wearable devices within the broader IoT ecosystem. Recent advances in consumer-grade Large Language Models (LLMs) have further simplified the development of wearable IoT applications by assisting with software development, system integration, and intelligent data processing. There are many commercial wearable headsets which are used in various applications, including opticians to diagnose various issues \cite{apple2024visionproavailability, sipatchin2021eyetracking}. There are also many other applications of wearable devices in Smart Healthcare \cite{Sundaravadivel_ICCE_2020} as shown in Fig. \ref{FIG:IoT}.   

% With high performance low power consuming devices available as off-the-shelf components, Internet of Medical Things (IoMT) attracted the attention of the world \cite{Sundaravadivel_ICCE_2020}. Fig. \ref{FIG:IoT} shows the various wearable devices. These devices has attracted the consumer market and the sales of wearables have increased significantly over the past few years. There are many uses of the wearable devices including, Virtual Reality, Medical, Military, and so on. 

This paper presents a PUF architecture that can be used to secure the resource-constrained wearable technologies. Many vulnerabilities and security concerns were identified in the software and hardware developed using LLMs. Georgia Institute of Technology researchers recently released an article showing the risks of developing the software and hardware using AI models \cite{gatech2026badvibes}. Google Threat Intelligence Group (GTIG) recently published an article showing the first known threat actor using a zero-day exploit believed to have been developed with AI \cite{gtig2026aithreattracker}. The vulnerability allowed bypassing two-factor authentication after valid credentials were obtained and that it was a semantic logic flaw involving incorrect developer assumptions. Currently, AI primarily enhances the capabilities of human attackers rather than replacing them. It accelerates research, software development, reconnaissance, phishing campaigns, and exploit development. \cite{gtig2026aithreattracker}. In this context, PUFs provide a hardware-assisted security primitive capable of generating device-unique cryptographic keys and strengthening hardware-rooted trust, thereby helping to mitigate several classes of attacks against resource-constrained IoT systems.

This paper is organized as follows:
Section \ref{SEC:Hardware_Assisted_Security} presents the state of the art of Hardware Assisted Security while the novel contributions of the paper are presented in Section \ref{SEC:novel_contributions}. An introduction to PUF, its working and applications are presented in Section \ref{SEC:PUF}. The architectures of PUF are presented in Section \ref{SEC:Architectures}. Section \ref{SEC:results} presents the simulation results and the conclusion and the future directions of this research is presented in Section \ref{SEC:conclusion}.

%%%%%%%%%%%%%%%%%%%%%%%%%%%%%%%%%%%%%%%%%%%%%%%%%%%%%%%%%%%%%%%%%%%%%%%%%%%%%%%%%%%%%%
\section{Security of Wearable Devices - Challenges and Hardware-Assisted Security (HAS) Approach as a Solution}
\label{SEC:Hardware_Assisted_Security}

Wearable devices have become increasingly prevalent in recent years \cite{ForbesWearable}. The global adoption of smartwatches and other wearable technologies continues to grow annually, driven by advances in sensing, connectivity, and artificial intelligence. These devices continuously collect physiological and environmental data, which are typically transmitted to cloud platforms for storage, analysis, and remote access. Wearable technologies have found widespread applications in both civilian and military domains, including health monitoring, fitness tracking, remote patient care, situational awareness, and mission-critical operations.

\subsection{Security Challenges of Wearable Devices}

%\textbf{\textcolor{red}{Goes here ...\\
%Write after going through and citing works like: \cite{Shrestha_ACMCS_2018-Jan}}}

Wearable devices offer numerous benefits that enhance consumers' quality of life. Fitness tracking is one of the key features that has contributed to the growing popularity of wearable technologies in the consumer market \cite{Sundaravadivel_ICCE_2020}. Researchers around the world are using the wearable devices for developing IoT architectures for applications such as sleep monitoring, thyroid monitoring \cite{Prabha_Thyroid} and other healthcare services. They constantly gather user information and transmit to the cloud. In certain applications, healthcare providers can remotely monitor patient data, enabling improved access to medical services and real-time health assessment \cite{Sundaravadivel_ICCE_2020}. 

\begin{figure}[htbp]
	\centering
	\includegraphics[width=0.72\textwidth]{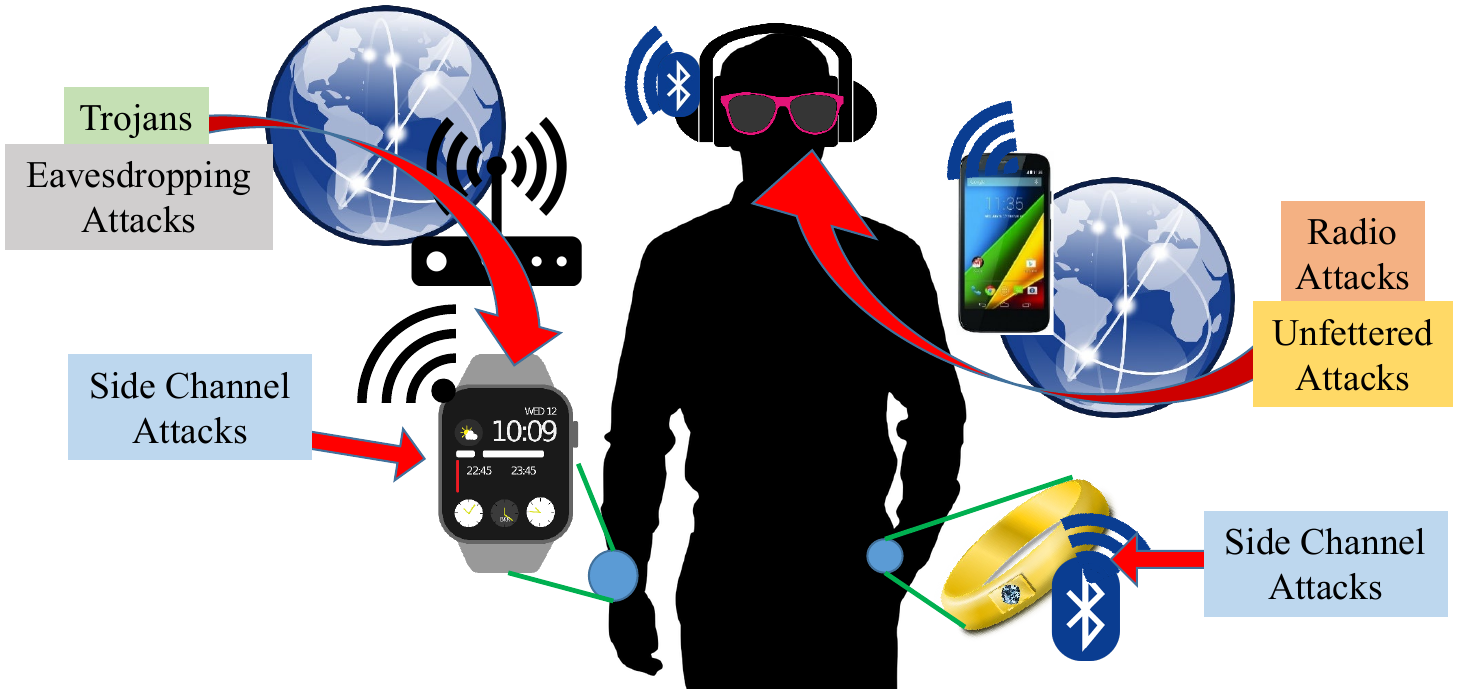}
	\caption{Security Challenges of Wearable Devices.}
	\label{FIG:Security_Challenges}
\end{figure}

Despite these advantages, wearable devices introduce security and privacy vulnerabilities that may expose sensitive user information to unauthorized entities \cite{Shrestha_ACMCS_2018-Jan}. Several commercial wearable devices have faced restrictions or bans in certain workplaces due to privacy concerns. For example, smart glasses can potentially record images or videos without the awareness or consent of individuals, creating risks in sensitive workplace environments. Additionally, wearable devices may collect personal information, including heart rate, location, and sleep patterns, when users enable such data-sharing features. Unauthorized access to this information could reveal sensitive details about users' health, daily activities, and lifestyles. Fig. \ref{FIG:Security_Challenges} illustrates the major security challenges associated with wearable devices.

There are many security vulnerabilities that affect wearable devices as shown in the figure \cite{Ahuja_IEEEECL_2020,Guo_IEEEIoT_2020,Dong_IEEEAccess_2019}. An attacker can get information from the devices using side channel attacks \cite{Liu_IEEETMC_2019}. These attacks take advantage of the vulnerabilities in the hardware rather than in the software. Attackers can exploit devices through side-channel attacks by analyzing power consumption, signal timing, and, in some cases, sound. Attackers also introduce Trojans into the devices and eavesdrop into the conversation in the network.

\subsection{Hardware-Assisted Security (HAS) - Brief Introduction}

%\textbf{\textcolor{red}{Goes here ...\\}}

Hardware-Assisted Security (HAS) is a promising approach for addressing these security challenges. A HAS is an extra module dedicated for security in the system. Numerous HAS solutions have been proposed for IoT applications, offering various levels of security capabilities. Some HAS modules can perform end-to-end encryption of the communication channel, while others do not have this capability \cite{Yanambaka_IEEETSM_2018, Yanambaka_2025_IEEEISVLI,aarella2025fortifiededge}. The functionality and design of a HAS module depend on the specific application requirements, system constraints, and security specifications.

One of the widely used HAS is Trusted Platform Module (TPM). It is an add-on module that has wide range of applications, from personal computers and servers to small IoT devices \cite{Shrestha_ACMCS_2018-Jan,Kim_CYSRAM_2019,Sahay_IEEETCE_2026}. A TPM is a multifunction module that can perform tasks like encryption, decryption, and other cryptographic operations like hashing. TPM can also produce random numbers for keys and in some cases like RSA generate public and private key pairs. One disadvantage is that unlike a PUF, a TPM produces pseudorandom numbers using a program to generate the random numbers.

Another commonly used HAS module is ARM Trustzone \cite{ArmTrustzone,Ma_IEEETIFS_2025}. This is a commercially available solution by ARM that is similar to a TPM but also provides a quarantine zone for performing the cryptographic tasks \cite{Yang_CUIC_2014}. Arm Trustzone has the ability to generate keys, perform encryption and decryption and provide hashing algorithms similar to a TPM. Arm also uses hardware based key generation similar to the PUF. However, one disadvantaeg of ARM TrustZone is that it is more expensive compared to a TPM.

The third widely used HAS module is PUF \cite{Yanambaka_CEM_2020,Yanambaka_2025_IEEEISVLI}. It can generate keys necessary for end-to-end encryption of communication channel, device identification and authentication in a network. They have several other applications as well. PUFs exploit naturally occuring nanoscale manufacturing variations to generate keys; therefore, the resulting keys are inherently random adding an extra layer of security to the overall system. \cite{Sahoo_IEEETC_2018}. Attackers cannot identify the keys generated by PUF module. A PUF on its own cannot perform cryptographic tasks, such as encryption and decryption, but when integrated with a cryptoprocessor, it can provide better security.

\subsection{Hardware-Assisted Security (HAS) is Best Suitable for Wearable Devices}

Wearable smart devices are embedded systems that run on batteries. These devices continuously monitor the user information and transmit to the cloud or store it locally for later access. This introduces privacy and security risks as discussed earlier. A misplaced wearable device, with a weak password can potentially put the user privacy at risk. 

Many software-based security solutions require high processing power or utilize significant device resources. Wearable devices have limited resources and must operate in a resource-constrained environment. Hence, adding an extra HAS module can help in the case of a wearable device. All the cryptographic processes will be off-loaded to the HAS which reduces the load on the main embedded processor.

A PUF module has the ability to generate multiple keys. With an appropriate PUF integrated into the design, it can generate multiple keys, potentially thousands in some architectures. This can be very helpful when the system is under attack. Although a PUF can generate a large number of keys, not all of them can be used for cryptographic purposes. Even when the system is under attack, the keys can be changed on-the-fly if necessary. 

%One of the advantages of HAS is it is an extra add-on module to the hardware. The processing is offloaded to the HAS module.
%\textbf{\textcolor{red}{Goes here ...\\}}

%%%%%%%%%%%%%%%%%%%%%%%%%%%%%%%%%%%%%%%%%%%%%%%%%%%%%%%%%%%%%%%%%%%%%%%%%%%%%%%%%%%%%
\section{Novel Contributions}
\label{SEC:novel_contributions}

\subsection{Problem Definition}
Previous sections discussed various security vulnerabilities associated with wearable devices. There are many vulnerabilities that an attacker can utilize and gain access to the system \cite{Ahuja_IEEEECL_2020,Guo_IEEEIoT_2020,Dong_IEEEAccess_2019,Ma_IEEETIFS_2025}. The ``\textit{things}'' in the IoT network are low-power resource-constrained devices that cannot execute resource-intensive cryptographic algorithms. Furthermore, the limited storage capacity of these devices makes it challenging to securely store long cryptographic keys required for end-to-end encryption. A device key, once compromised, can potentially provide access to the entire IoT ecosystem. 

\subsection{Proposed Solution}
HAS is one of the potential solutions for IoT and wearable devices. PUF modules help achieve robust security in IoT environments. Keys generated by a PUF module cannot be reproduced in any other module. Even if the attacker gets hold of the key, it can be changed on-the-fly using an appropriate PUF architecture.

This paper analyzes two designs of PUF architectures, Arbiter and Hybrid Oscillator Arbiter (HOA) PUF. These two PUF Architectures were selected for IoT security because they exhibit a very low key generation error rate. The key size can be scaled according to the available resources, and the proposed designs are reconfigurable. This paper presents the PUF architectures, their simulation results and an analysis demonstrating their sustainability for tailored IoT architectures.

%%%%%%%%%%%%%%%%%%%%%%%%%%%%%%%%%%%%%%%%%%%%%%%%%%%%%%%%%%%%%%%%%%%%%%%%%%%%%%%%%%%%%%%%%%%%%%%%%%%%%%%%%%%%%%%%%%%%%%%%%%%%%%%%%%%%%%%%%%%%%%%%%%%%%%%%%%%%%%%%%%%%%%%%%%%%%%%%%%%%%%%%%%%%%%%%%%%%%%%%%%%%%%%%%%%%%%%%%%%%	
\section{State Of The Art Of Physical Unclonable Function}
\label{SEC:PUF}

During the fabrication of an Integrated Circuit (IC), several manufacturing variations get introduced that are uncontrollable, unavoidable, unpredictable and naturally occurring. Although technology has advanced sufficiently to manufacture ultra low power devices at very small technology nodes, the manufacturing variations have been affecting the yield of the final chips for decades \cite{Zheng_2014_IEEETED}. Moreover, they not only degrade device performance but also influence aging effects throughout the device's operational lifetime \cite{Li_IEEEJEDS_2025}. 

% A PUF takes in challenges as inputs and gives the responses as outputs \cite{Yanambaka_CEM_2020}. The steps involved during the fabrication of an Integrated Circuit (IC) introduces nanoscale variations into the devices. These affect the functioning of the devices. The manufacturing nanoscale variations are unpredictable, unavoidable, uncontrollable and natural. When the circuits are designed fabricated on the wafer with these naturally occurring variations in place, no two devices look alike. Two identical designs made from these devices do not produce identical outputs. PUF modules, designed with these variations, gives outputs which act as fingerprint for the respective IC. The working principle of PUF is shown in Fig. \ref{FIG:WorkingOfPUF}. 

\begin{figure}[htbp]
	\centering
	\subfigure[Single PUF with multiple inputs gives multiple outputs.]{\label{fig:PUFsingle}\includegraphics[width=0.45\textwidth]{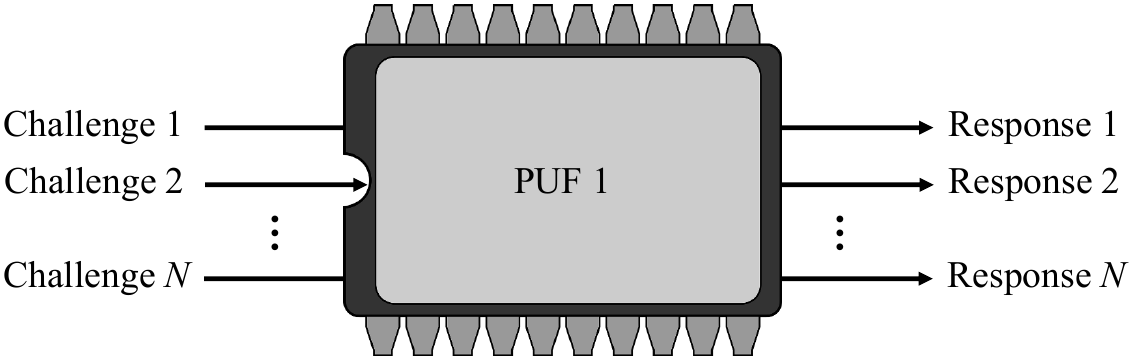}}
	\subfigure[Multiple PUFs with same input gives multiples outputs]{\label{fig:PUFmultiple}\includegraphics[width=0.5\textwidth]{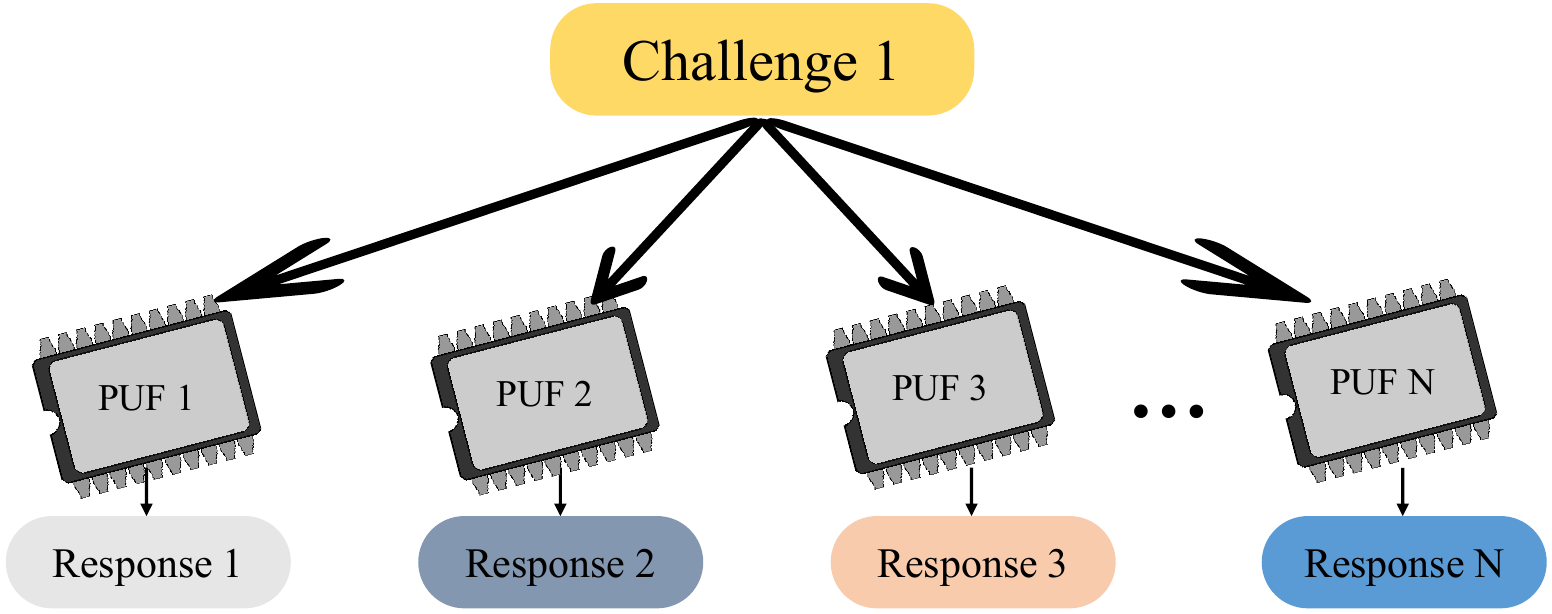}}
	\caption{Working of a PUF.}
	\label{FIG:WorkingOfPUF}
\end{figure}

PUF uses these manufacturing variations to generate the random numbers \cite{Yanambaka_2025_IEEEISVLI}. These random numbers generated by the PUF are used for a wide range of cryptographic applications, including smart healthcare, smart agriculture, smart cities, and other IoT applications \cite{aarella2025fortifiededge, sadhu2024easysec}. Fig. \ref{FIG:WorkingOfPUF} shows the working of a PUF. Consider two PUF modules, $P_1$ and $P_2$. The input to a PUF is called a ``challenge" and the random number output generated from the PUF module is called ``response''. Given a PUF module $P_1$, a challenge produces a corresponding response and these two are called ``challenge reponse pair'' (CRP) for that module. Every new challenge of PUF must produce a new response. If ``N'' challenges are given to a PUF, it generates ``N'' responses.

PUFs have numerous cryptographic applications, including device authentication, encryption, blockchain, and particularly in
mission-critical domains such as healthcare, defense, and military systems. To employ PUF modules in cryptographic applications, the PUF modules must satisfy Figures of Merit (FoM) which are discussed further in section \ref{SEC:results}.
% As shown in Fig. \ref{fig:PUFsingle}, if `\textit{N}' challenges are given to a PUF, it generates `\textit{N}' unique responses that satisfy the FoMs. If the same challenge `$C_1$' is given to `\textit{N}' PUF modules, they must generate `\textit{N}' unique responses that satisfy the FoMs. 

% The input to the PUF is a ``Challenge'' and the output from the PUF module is called ``Response''. One set of input and output is called ``Challenge Response Pair'' (CRP). With a new challenge, the PUF gives out a new response. Sections \ref{SEC:Arbiter_PUF}, \ref{SEC:HOA_PUF} presents a deep analysis of the Arbiter and Hybrid Oscillator Arbiter PUF which discusses more on the working of a PUF. As shown in Fig. \ref{fig:PUFsingle}, when multiple challenges are given to a PUF module, multiple responses are generated. Considering multiple PUF modules, if same challenge is given to all the modules, they produce multiple responses. 

% One of the applications of PUF is cryptography. The output responses to be used for cryptographic applications, there are certain properties to be satisfied by the PUF responses which will be discussed in detail in Section \ref{SEC:results}. All the responses generated by PUF cannot be used for all applications. All PUFs are not able to generate multiple PUF responses that can be used for various applications. Based on the number of challenge inputs and responses that can be used on a PUF, four types of PUF architectures are used widely \cite{Emerging_PUF_IEEEaccess}: 

PUFs are classified into various architectures based on their underlying core components. These architectures include SRAM PUFs, Ring Oscillator (RO) PUFs, Quantum PUFs, and other variants. \cite{Bathalapalli_IoTJ_2025, Aging_ROPUF, Yanambaka_2025_IEEEISVLI}. PUFs are classified into four categories based on the number of keys each architecture can generate while satisfying the required FoMs \cite{Emerging_PUF_IEEEaccess} as shown in Fig. \ref{FIG:PUF_Types_CRP}: 

\begin{figure}[hb]
	\centering
	\includegraphics[width=0.70\textwidth]{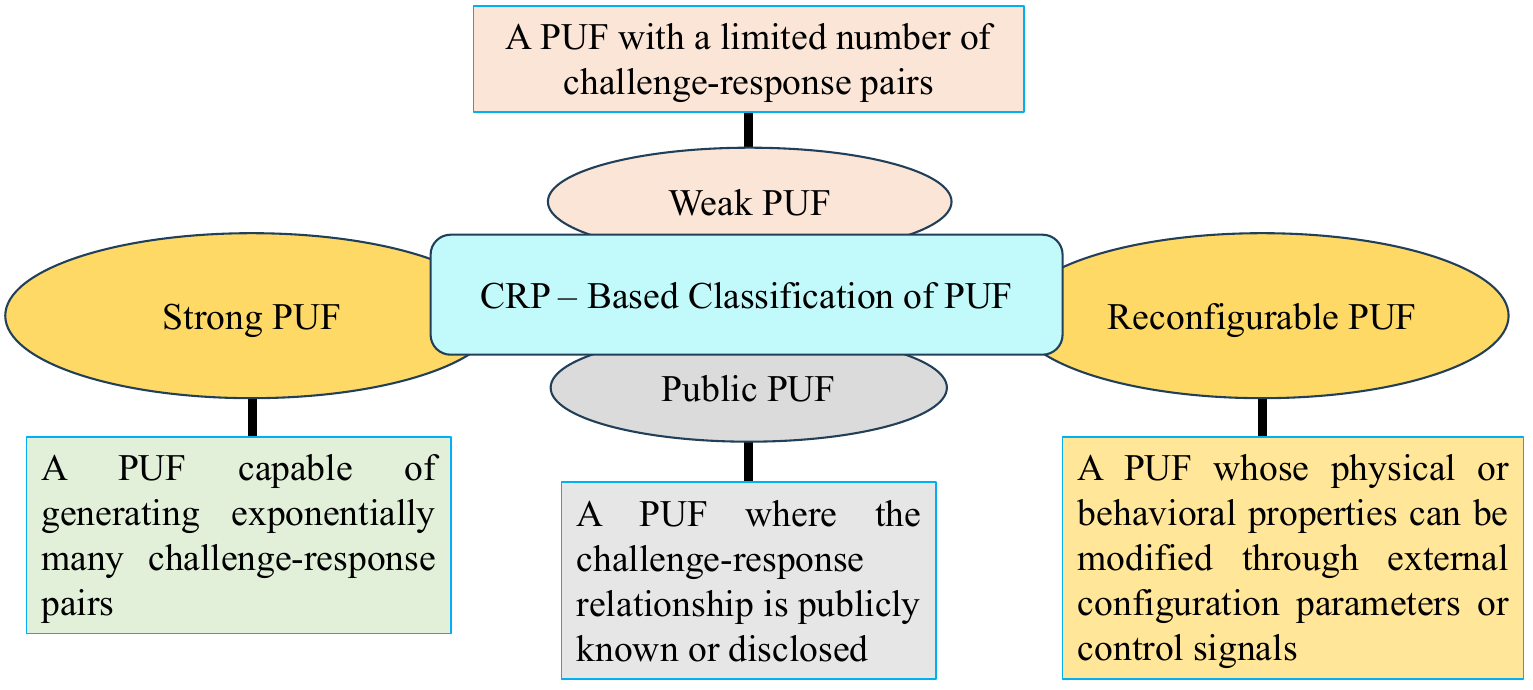}
	\caption{CRP - Based Classfication of PUF.}
	\label{FIG:PUF_Types_CRP}
\end{figure}

\begin{itemize}
	\item Weak PUF
	\item Strong PUF
	\item Reconfigurable PUF
	\item Public PUF
\end{itemize}

\subsection{Weak PUF}
\label{sec:Weak_PUF}
Weak PUFs can generate only a limited number of random numbers that satisfy the Figures of Merit (FoMs). In certain cases, the number can be as low as '1'. The number of keys generated that can satisfy the FoMs depends on the manufacturing variations present in the IC. One such architecture that generates limited number of reliable cryptographic keys is SRAM PUF \cite{Pratihar_IEEETCAD_2024}. One limitation of SRAM PUF that results in a low number of usable keys is the occurrence of errors during key generation. Various factors, such as temperature, voltage fluctuations, and humidity, affect its performance, necessitating the use of fuzzy extractors. Although it is a weak PUF, almost every IoT device or board has an SRAM module, making its implementation relatively easier.

% PUF modules cannot generate multiple responses that can satisfy all the properties required for cryptographic applications. In some architectures, out of all the PUF responses generated, very low number of them can be used for cryptographic applications. In some cases, that number can be as low as 1. Only that one response can be used for identification or encryption purposes. Those architectures are called Weak PUFs. Example of weak PUF is SRAM PUF architectures \cite{SRAM_PUF_2009}. In SRAM architectures, the outputs are also noisy which requires additional fuzzy extractors to generate a reliable response. 

\subsection{Strong PUF}
\label{sec:Strong_PUF}

Some PUF architectures, such as Arbiter PUF, can generate a larger number of cryptographic keys that satisfy the FoMs. These are called "Strong PUFs" \cite{Yanambaka_CEM_2020}. This increases the robustness of the PUF and expands its range of applications. In certain cases, the number of usable keys can reach several hundreds. Unlike SRAM PUF, most of these architectures are application-specific modules requiring additional hardware on the chip, resulting in trade-offs between battery life and performance. Recent developments in quantum technology have also led to the emergence of quantum PUF architectures.

% In some architectures of PUF, most of the responses generated can satisfy all the properties of PUF and can be used for cryptographic purposes. Such architectures are called Strong PUFs. This gives an advantage to generate multiple keys and an increased robustness in security. Example of such strong architectures is Arbiter PUF \cite{inis_prasanth_2016}. The number of responses from a strong PUF can be in hundreds or thousands in certain architectures. 

\subsection{Reconfigurable PUF}
\label{sec:Reconfigurable_PUF}

Architectures such as Arbiter PUFs remain vulnerable to machine learning attacks \cite{Yanambaka_2025_IEEEISVLI}. With the rapid growth of AI, even the natural randomness introduced in PUF modules is susceptible to modeling attacks. To resist such adversaries, PUFs can be reconfigured on-the-fly to generate multiple responses. This capability proves particularly valuable in applications requiring a high volume of keys, such as networking.

% Strong PUF generates multiple PUF responses that can be used for cryptographic applications. But in some architectures, the PUF responses are vulnerable to machine learning attacks. PUFs like Arbiter PUF are vulnerable to machine learning attacks. Hence reconfigurable architectures are introduced in PUF designs. This gives an advantage to reconfigure the PUF on-the-fly and generate multiple responses. This significantly increases the number of responses from a single PUF. In some applications such as networking, the number of responses required can be very high. These reconfigurable PUFs can be used in such applications for high robustness and security. 

\subsection{Public PUF}
\label{sec:Public_PUF}

A public PUF differs fundamentally from the traditional PUF. Unlike Strong or Weak PUFs, where the challenge-response behavior must remain confidential to preserve security, a Public PUF is characterized by having a publicly available mathematical model that can simulate its behavior. However, the critical distinguishing feature is that while this model can accurately predict the PUF's responses, doing so requires significantly more computational time and resources than physically querying the actual device itself. This asymmetry between the ease of physical measurement and the computational difficulty of simulation forms the security foundation of Public PUFs. This property enables unique cryptographic protocols, such as certificate-based authentication and secure key exchange \cite{Emerging_PUF_IEEEaccess}.

% When a PUF is deployed in the applications, some require the architecture to be hidden from the user or consumers. In public PUF architectures, the architecture is known to the user. In such cases the reconfigurability can be highly useful as it generates significantly high number of responses. 

\subsection{Other PUF Classifications}
\label{sec:Other_Classifications}

The classification mentioned above is based on the number of CRPs generated that can satisfy the FoMs. PUF is classified into two categories based on the material used to design the PUF. They are Silicon PUFs and Non Silicon PUFs. 

Silicon PUFs are the most widely studied and implemented category, leveraging inherent manufacturing variations present in silicon-based semiconductor devices such as transistors, gate delays, and memory cells. These variations arise naturally during the CMOS fabrication process due to microscopic differences in doping concentrations, oxide thickness, and threshold voltages, even when devices are manufactured using identical masks and processes. The Arbiter PUF, SRAM PUF, and Ring Oscillator (RO) PUF architectures discussed in
the previous subsections are examples of silicon-based PUFs. \cite{Pratihar_IEEETCAD_2024,Sahoo_IEEETC_2018,Yanambaka_iNIS_2016_Multi-Key-PUF}. 

Non-Silicon PUFs, on the other hand, derive their unique responses from physical phenomena outside conventional silicon-based transistor structures. These architectures exploit properties such as optical scattering, magnetic characteristics, or intrinsic material variations to generate unclonable identifiers. One such architecture is the Optical PUFs, which utilizes the unique scattering patterns produced when light passes through randomly structured optical media \cite{Lu_OpticalPUF_2018}. The other type of Non-Silicon PUFs is the Magnetic PUFs, which rely on the unpredictable magnetic domain patterns in materials \cite{Wang_IEEETIM_2026}. With the advancements in Quantum technology, quantum-based PUFs have also been introduced for post-quantum cryptographic applications \cite{Bathalapalli_IoTJ_2025}.

\begin{figure}[htbp]
	\centering
	\includegraphics[width=0.75\textwidth]{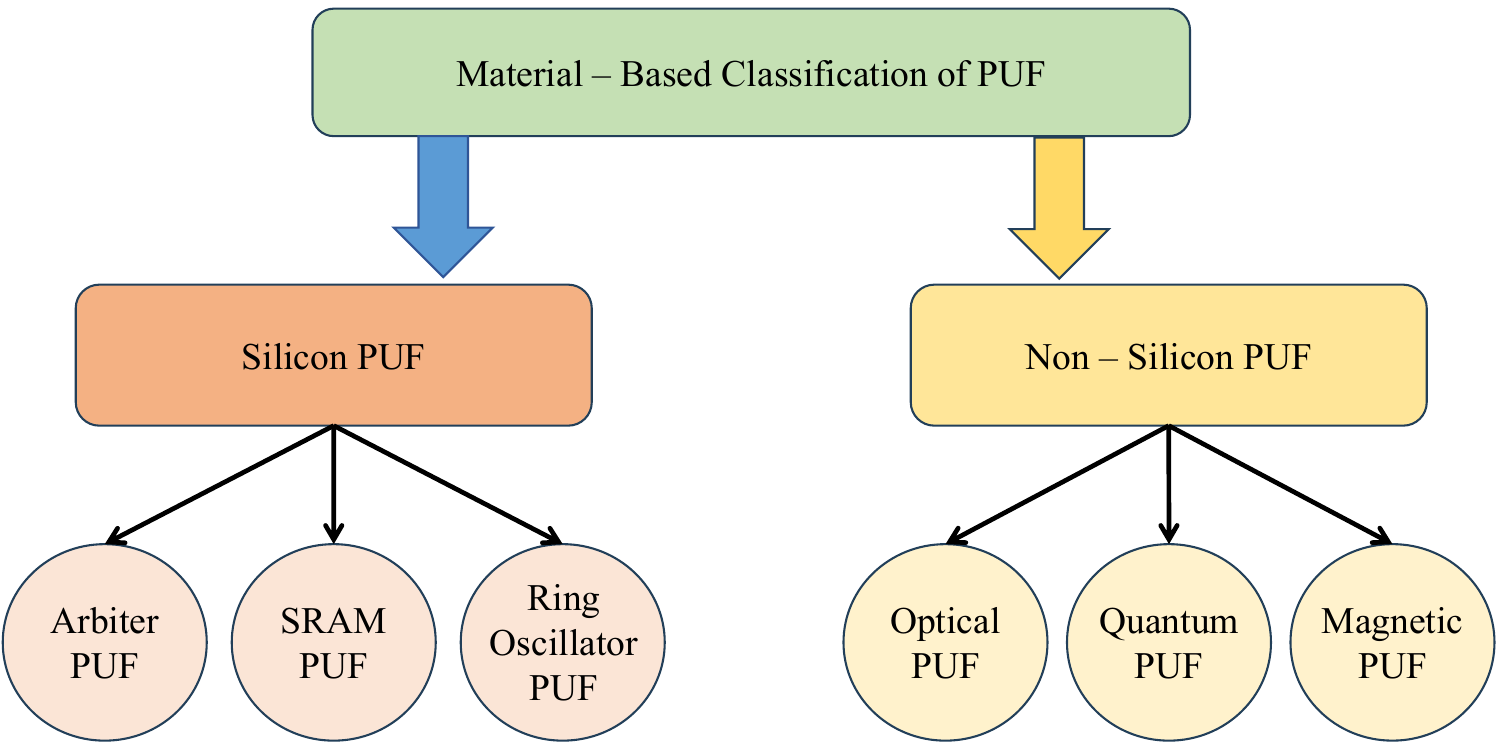}
	\caption{Material - Based Classfication of PUF.}
	\label{FIG:PUF_Types_Material}
\end{figure}

\section{Related Prior Research}
\label{SEC:related_research}
	
%\begin{table}[htbp]
%	\centering
%	\caption{Currently Available Security Protocols for Wearable Devices.}
%	\label{Table:Current_Wearable_Security}
%	\begin{tabular}{|c|c|c|c|}
%		\hline
%		Research Work & Anonymity & Visibility & Power Consumption\\
%		\hline
%		
%	\end{tabular}
%\end{table}

%\begin{table*}[htbp]
%	\centering
%	\caption{Currently Available Security Protocols for Wearable Devices.}
%	\label{Table:Current_Wearable_Security}
%	\begin{tabular}{|c|c|c|c|}
%		\hline
%		Wearable Device & Application & Security Vulnerabilities & Security Solutions\\
%		\hline
%		Wrist Worn& Wrist worn devices with interactive display, Tracking Devices & &\\
%		\hline
%		
%	\end{tabular}
%\end{table*}

%\begin{table*}[htbp]
%	\centering
%	\caption{Currently Available Security Protocols for Wearable Devices.}
%	\label{Table:Current_Wearable_Security}
%	\begin{tabular}{|c|p{3.5cm}|p{3.5cm}|p{1.5cm}|p{1.5cm}|p{1.1cm}|p{1.0cm}|}
%		\hline
%		Research Work &  Application & Network Connection Required & Cost Effective & Ease of Use & Anonymity & Visibility\\
%		\hline
%		Yamada et. al \cite{Yamada_CMS_2013} & Protection against unintentional face image capture & -- & -- & $\checkmark$ & $\checkmark$ & $\checkmark$\\
%		\hline
%	\end{tabular}
%\end{table*}

\begin{table*}[htbp]
	\centering
	\caption{Currently Available Security Protocols for Wearable Devices.}
	\label{Table:Current_Wearable_Security}
	\begin{tabular}{|p{0.12\textwidth}|p{0.38\textwidth}|p{0.38\textwidth}|}
		\hline
		\textbf{Research Works} & \textbf{Security Application} & \textbf{Specific Properties} \\
		\hline
		\hline
		Yamada et. al \cite{Yamada_CMS_2013} & Protection against unintentional face image capture & \begin{itemize} [leftmargin=*, topsep=0pt, itemsep=0pt, parsep=0pt]
			\item A dedicated device is required for the proposed work
			\item Easily accessible by user
			\item Anonymity of user.\end{itemize} 
	\\
		\hline
		Henne et. al \cite{Henne_ACM_2013} & Prevent infringing media through geo-tagged media & \begin{itemize} [leftmargin=*, topsep=0pt, itemsep=0pt, parsep=0pt]
			\item Can be used in Smart Phone
			\item Network connection is required
			\item Cost effective
			Third party services required\end{itemize}
	\\
	\hline
	Hoyle et. al \cite{Hoyle_ACM_2014} & Lifelogging cameras were used for exploring privacy management of people in daily life & \begin{itemize} [leftmargin=*, topsep=0pt, itemsep=0pt, parsep=0pt]
		\item Cameras were required 
		\item Does not require network connectivity on wearable
		\item Privacy of the user and others around them were tested \end{itemize}
\\
	\hline
	Ashok et. al \cite{Ashok_ACM_2014} & Beacons in user eyewear communicate with camera to select privacy preferences & \begin{itemize} [leftmargin=*, topsep=0pt, itemsep=0pt, parsep=0pt]
		\item Embedded into user eyewear
		\item No connectivity is required
		\item User interaction is not necessary
		\item Uses IR signals to transmit the privacy settings \end{itemize}
\\
	\hline
	\end{tabular}
\end{table*}

With the rapid advancement of AI, the demand for high performance and high throughput devices has increased significantly.
In particular, the memory bandwidth of the GPU has increased by more than tenfold over the past decade \cite{nvidia_1080,nvidia2024h200}. There are several challenges associated with implementing AI at the scale required \cite{Kusnezov_HCS_2021}. Among these, the power consumption and the performance of the transistor devices are the two most significant challenges. In particular, research on transistor devices focuses on device scaling with an intent to maximize density, high throughput and ultra-low power consumption.

Technology scaling has reached the atomic level \cite{bean2026intel18a}. Technology scaling has reached a level that has never been seen before. Such atomic scaling levels were not possible with the planar transistors. Bulk CMOS transistors could scale well until the 32nm technology node. However, the one dimensional planar CMOS transistor geometry was not sufficient for scaling beyond 32 nm into the 22nm. Consequently, 3-dimensional FinFETs were developed to overcome this scaling challenge of conventional planar transistors \cite{iet_2016, AMD}. Due to this increase in transistor density on a single die, high performance SoCs and NoCs have become feasible which accelerated growth and widespread development of IoT.

% Aim of the research and development of devices and even scaling of transistors is to achieve highest possible throughput with minimum power consumption possible. The current technology has already reached a 10nm range \cite{snapdragon835}. FinFETs were developed for scaling in the first place to achieve low power consumption compared to the planar transistors. A comparative analysis of FinFETs and the planar transistors is presented in \cite{iet_2016}. Due to this increase in transistor density on a single die, high performance SoCs and NoCs have become possible which opened gates to the IoT development.

The price of the final chip is decided based on the yield \cite{Mohanty_Book_2015_Mixed-Signal}. The yield in particular is affected by the manufacturing variations introduced during the fabrication of the IC. As mentioned in previous section, they are naturally occurring, unavoidable, unpredictable and uncontrollable. Hence the research community started using the naturally introduced variations for cryptographic purposes. These are used for generating the natural random numbers using the PUF modules. 

% Research that has been conducted through the ages has been to decrease the number of problems during design and manufacturing. Even in the semiconductor industry, the yield decides the price of the final IC \cite{Mohanty_Book_2015_Mixed-Signal}. But the nanoscale manufacturing variations cannot be avoided. As mentioned in previous section, they are naturally occurring, unavoidable, unpredictable and uncontrollable. Hence the research community started using the naturally introduced variations for cryptographic purposes. This adds natural randomness to the PUF modules which is reliable compared to the psuedorandomness in certain cases and in some others, necessary. 	

% This has been possible due to the research advancements in various areas such as device nodes, communication and processor throughput. The power consumption of devices has reduced with introduction of each technological node. Bulk CMOS has been around until the devices reached 32 nm. Transistors smaller than 32 nm followed a different design, FinFETs. It has been more than a decade since FinFETs 22 nm node was introduced in commercial applications and now-a-days the commercial technological node has reached 7 nm \cite{AMD}. When a new technology node has been introduced, it takes some time for the fabrication to yield  maximum working ICs with minimal nanoscale manufacturing variations. So PUF started gaining more visibility in the research community.  

With the  availability of high performance and low power technology nodes, researchers around the world have started developing IoT architectures for a wide range of applications, including smart healthcare, wearable technologies, and smart transportation over the last few decades. For example, a quadrature design using only standard components was developed in 2016 by Kougianos et al. for real-time object tracking \cite{Kougianos_IEEEAccess_2016}. A thyroid monitoring system was proposed by Prabha et al. in 2016 \cite{Prabha_Thyroid} as an application of smart healthcare. Recently, commercial wearable virtual reality (VR) headsets have become widely available, demonstrating the rapid evolution and growing maturity of wearable technologies \cite{apple2024visionproavailability}. The availability of highly capable commercial AI platforms, such as OpenAI, Google Gemini, and Anthropic's language models, has simplified software development by assisting with code generation, debugging, documentation, and system integration \cite{openai2026gptlatest,anthropic2026fable5,google2026gemini31}.

Widespread adoption of AI models has increased cybersecurity challenges \cite{anthropic2026fable5statement}. Commercially available AI models
have brought to light several unknown software vulnerabilities, including zero-day vulnerabilities \cite{microsoft2026npmdependencyconfusion}. Furthermore, large language models (LLMs) and agentic AI systems have increasingly been used to discover security flaws in commercial and enterprise-grade software and hardware \cite{redhat2025libvirtprivesc}. A single solution might not be able to resist these attacks, but one of the solutions is to use Hardware Assisted Security (HAS). PUF is one such  HAS primitive that can generate true random numbers using manufacturing variations. Research based on PUF has been conducted extensively over the last few decades. In \cite{Suh_DAC_2007}, a ring oscillator-based PUF (RO PUF) was developed. In \cite{Haider_IoTPTS_2016}, a trusted sensor for mobile applications was proposed. Furthermore, in \cite{Mathew_ISSCC_2014} a PUF was presented that is tolerant to process variations, temperature changes, and voltage fluctuations.

This paper mainly discusses two architectures of PUF: RO PUF and Arbiter PUF \cite{Improved_ROPUF_2009,Arbiter_PUF}. These PUF
architectures have been evaluated and implemented in many applications. Several other PUF architectures have also been proposed, including SRAM PUFs \cite{SRAM_PUF_2009}, which have been studied since the early development of PUF-based security solutions.  One advantage of SRAM PUF is that most IoT devices already contain embedded SRAM, which can potentially be used to generate keys using PUF. However, SRAM PUFs are susceptible to environmental variations and noise during key reconstruction, necessitating the use of additional error correction mechanisms, such as fuzzy extractors, to reliably generate cryptographic keys.

An Oscillator Collapse Physical Unclonable Function (OC-PUF) was proposed by Zayed et al. in \cite{Zayed_IEEE_Access_2021}. The architecture proposed in the paper is focused mainly on the power consumption that is more suitable for devices with resource constraints. The authors used a 20nm trigate FinFET architecture for the simulation. The overall power consumption of the device is about 140 nW with the worst case simulations showing a 750 nW power consumption. The reliability of the proposed design is 99 \% with a uniqueness of 50.01 \%. 

A Hybrid Ring Oscillator Arbiter PUF is proposed by Driemeyer et al. in \cite{Driemeyer_IEEEJSSC_2026}. The authors utilized a 28nm planar CMOS technology to simulate and evaluate the proposed design. The primary objective of this architecture is to develop a high-speed PUF capable of generating cryptographic keys at a high bit rate while maintaining low power consumption. The PUF architecture in \cite{Driemeyer_IEEEJSSC_2026} generates the keys at a rate of 4.515 Gbits/sec while consuming about 0.259 pJ/bit. 

Pahlevi et al. proposed a Per-Selection-Enhanced Arbiter PUF architecture in \cite{Pahlevi_IEEEAccess_2025}. The authors evaluated the proposed design using 56 FPGA boards to assess its performance and reliability. The primary focus of the architecture presented in \cite{Pahlevi_IEEEAccess_2025} was improving the authentication performance, particularly in terms of the False Acceptance Rate (FAR) and the False Rejection Rate (FRR), which remained below or close to 2.5.

Das et al. proposed a PUF architecture based on voltage-gated spin-orbit torque magnetic tunnel junctions (VGSOT-MTJs) in \cite{Das_IEEEJN_2025}. The proposed PUF architecture achieves a uniqueness of 50.2\% and a reliability of 97.3\%. Furthermore, the architecture demonstrates strong resistance against machine learning (ML)-based modeling attacks. For example, a multilayer perceptron (MLP) attack achieved a prediction accuracy of less than 55.27\%, indicating improved resilience against ML-based attacks. The proposed PUF consumes approximately 63.67 fJ/bit while generating cryptographic keys at a bit rate of 0.27 Gb/s.

\section{PUF Architectures of Energy-Efficient Low-Overhead Wearable Security}
\label{SEC:Architectures}

This section presents the HOA-PUF architecture for energy-efficient low-overhead wearable security. The core components of the HOA-PUF are Ring Oscillator (RO) PUF and the Arbiter PUF. Both of these architectures have been extensively studied over the years and have demonstrated reliable performance. These PUF architectures can be implemented on an FPGA without requiring fabrication, making them suitable for rapid prototyping and performance evaluation in low-power applications, such as the Internet of Things (IoT). This provides researchers with the advantage of readily integrating the same module into IoT environments for rapid prototyping, testing, and deployment. 

% This section presents the PUF module architectures that are in question of a potential added security module for low power IoT devices. This section presents the Arbiter PUF and Hybrid Oscillator Arbiter (HOA) PUF architectures. The reason behind choosing these designs is the architectures has been around for quite some time and much research was conducted in this area. This qualifies the designs to be reconsidered for low power applications like IoT. Also, these are the two designs that can be implemented on an FPGA without fabrication. This gives researchers an added advantage for introducing the same module in their IoT environments. 

%%%%%%%%%%%%%%%%%%%%%%%%%%%%%%%%%%%%%%%%%%%%%%%%%%%
\subsection{Ring Oscillator PUF}
\label{sec:RO_PUF}
\begin{figure}[!ht]
	\centering
	\includegraphics[width=0.75\textwidth]{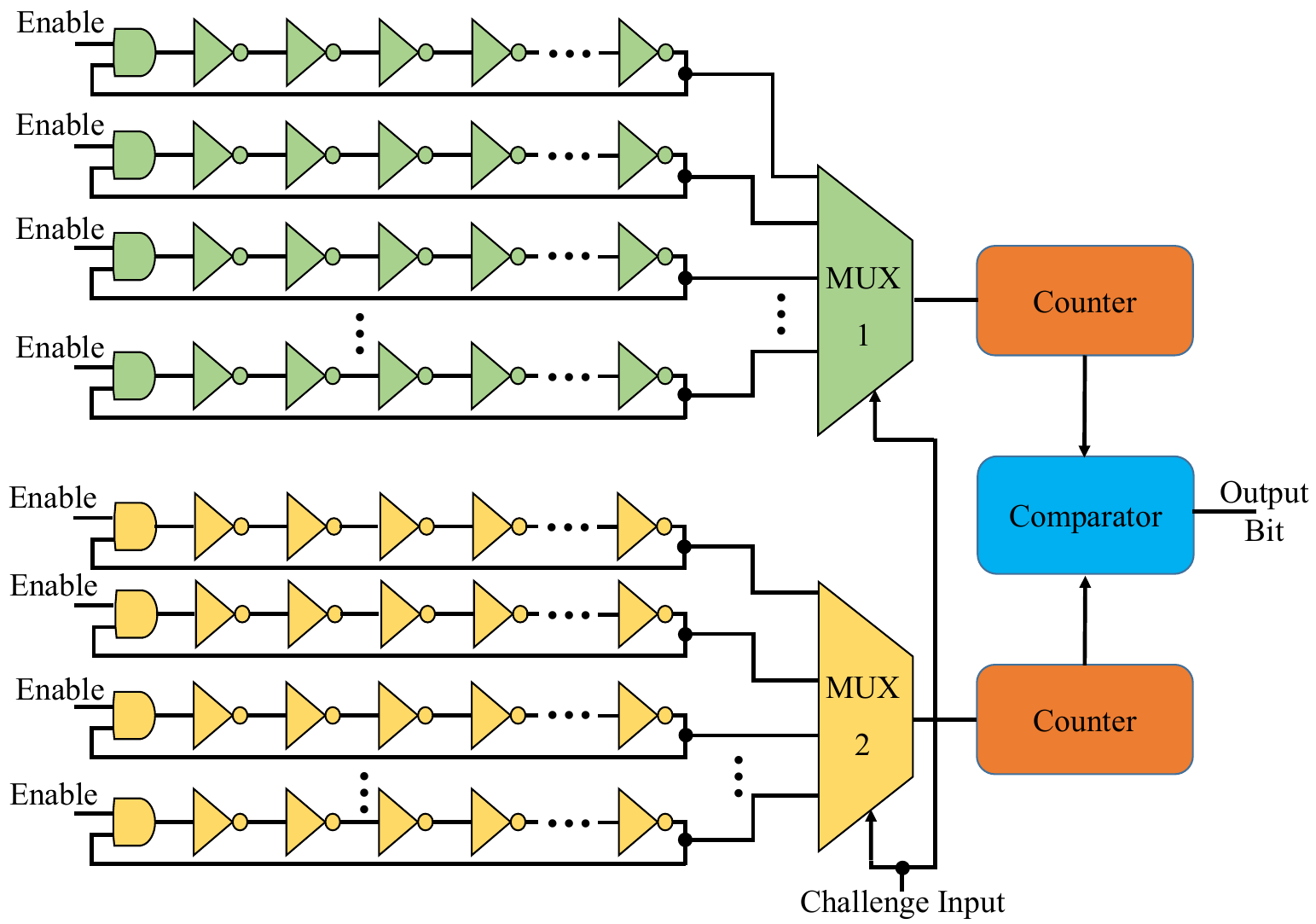}
	\caption{Ring Oscillator PUF.}
	\label{FIG:RO_PUF}
\end{figure}

The architecture of RO PUF is shown in Fig. \ref{FIG:RO_PUF}. The inverter is the fundamental building block of an RO PUF. As illustrated in the figure, the architecture consists of two sets of ROs connected to multiplexers. The challenge inputs are applied as select lines to the multiplexers. When a challenge is applied to the PUF, one RO is selected from each set and the resulting oscillation signals are passed to the counters. These counters measure the frequencies of the selected ROs. Finally, a comparator evaluates the frequency difference between the selected ROs and generates the output response bit as either '1' or '0'.  

The inverter, as the fundamental component of an RO PUF, captures inherent manufacturing variations, including gate delays, dopant concentration variations, and temperature-induced parameter shifts, which influence the final key generation process. Although all ring oscillators are designed identically with the same number of stages, process variations during fabrication introduce subtle, yet significant, differences in signal propagation delays. These variations contribute to the unique physical characteristics of each device and enable the generation of device-specific random responses.

% Fig. \ref{FIG:RO_PUF} shows the design of RO PUF. As shown in the figure, the core component of the RO PUF is the inverter. The RO PUF uses the nanoscale manufacturing variations to generate the keys. The delay added by the transistors in the inverters changes the oscillations.  As shown in the figure, the design has ring oscillators, multiplexers, counters and a comparator. The multiplexers select the oscillators one at a time and feed the oscillations to the counters.

% All the ring oscillators have identical design and identical number of inverters. The nanoscale manufacturing variations introduce errors and no two oscillators produce the same oscillation frequency. The challenge from the user is given to the multiplexers and oscillators are selected. The frequency of oscillations is calculated by the counters. The frequency is compared using the comparator and output bit is produced. Changing the challenge changes the selection of oscillators and the output key.

%%%%%%%%%%%%%%%%%%%%%%%%%%%%%%%%%%%%%%%%%%%%%%%%%%%%%%%%%%%%%%%%%%%%%%%%%%%%%%%%%%%%%%%%%%%%%%%%%%%%%%%%%%%%%%%%%%%%%%%%%%%%%%%%%%%%%%%%%%%%%%%%%%%%%%%%%%%%%%%%%%%%%%%%%%%%%%%%%%%%%%%%%%%%%%%%%%%%%%%%%%%%%%%%%%%%%%%%%%%%%%%%%%%%%%%%%%%%%%%%%%%%%%%%%%%%%%%%%%%%%%%%%%%%%%%%%%%%%%%%%%%%%%%%%%%%%%%%%

\subsection{Design of Arbiter PUF}
\label{SEC:Arbiter_PUF}
	
% Arbiter PUF uses the delay present in the devices to generate the PUF keys. Fig. \ref{FIG:MUX_Arbiter_PUF} shows the design of Arbiter PUF. As shown in the figure, the core component of the Arbiter PUF architecture is the Multuplexer. The nanoscale manufacturing variations introduced during the fabrication of the module introduces the delay. With many devices used for the Arbiter chain, the delay added gets accumulated makes it significant. 

\begin{figure}[!ht]
		\centering
 		\includegraphics[width=0.75\textwidth]{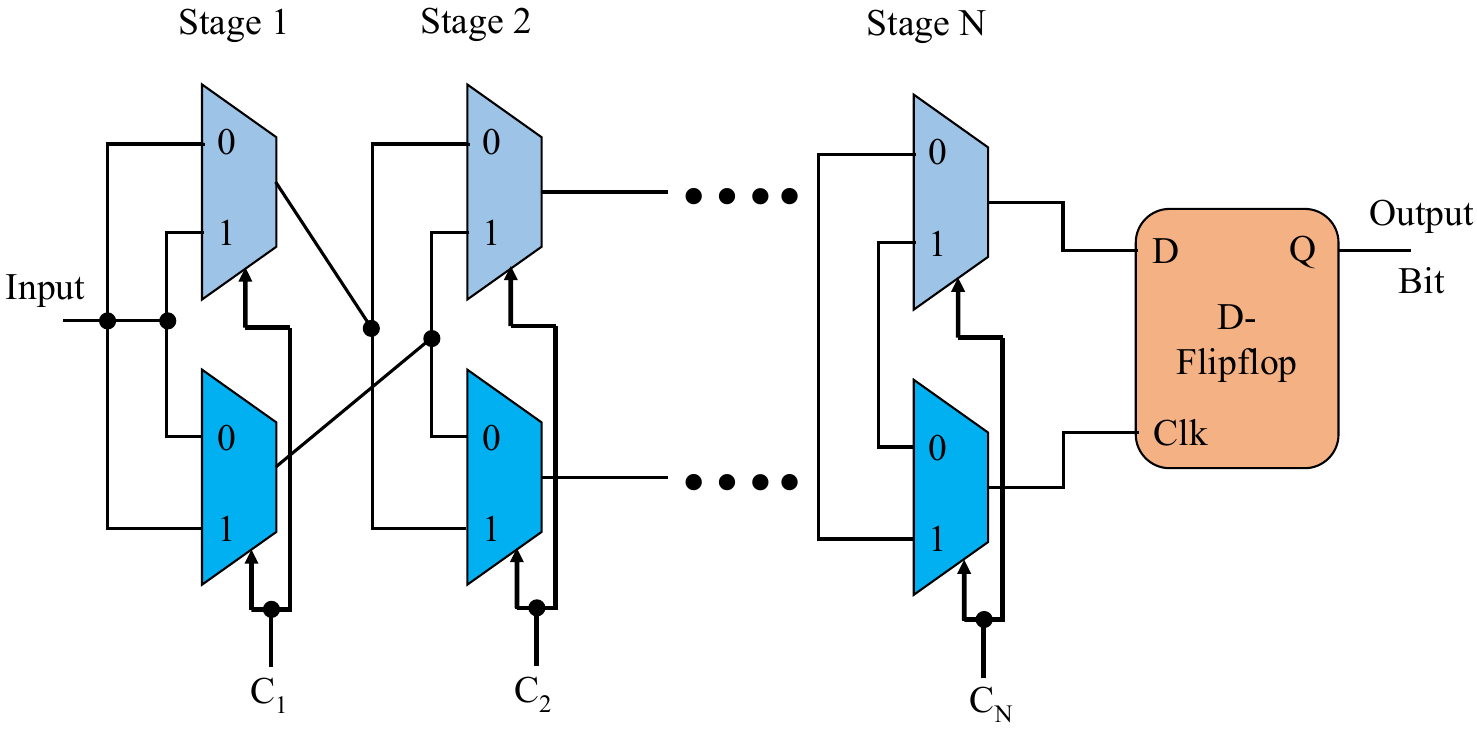}
		\caption{Design of Arbiter PUF.}
		\label{FIG:MUX_Arbiter_PUF}
\end{figure}

Fig.\ref{FIG:MUX_Arbiter_PUF} illustrates the architecture of the Arbiter PUF consisting of two series of multiplexers. The output of the multiplexers in the first stage is connected to the input of the multiplexers in the subsequent stage. The challenge given to the PUF feeds the select lines of the multiplexers. The outputs of the multiplexers in the final stage are connected to a D flip-flop, with one output connected to the D input and the other connected to the clock input of the flip-flop. The flip-flop determines the final response bit based on the relative delay difference between the two paths.
	
The core component of an Arbiter PUF is the multiplexer. These multiplexers capture manufacturing variations and subtle differences in transistor geometries introduced during the fabrication process. The propagation delay introduced by each multiplexer stage accumulates through the delay paths, ultimately affecting the timing relationship between the two signals arriving at the D flip-flop. An input is applied simultaneously at the first stage of both series of multiplexers. The path this input takes through both the series of Multiplexers differs based on the challenge given to the PUF. Due to the gate delay added by each transistor and other components, the signals reach the D-flipflop at different times. If the signal connected to the D input arrives before the clock signal, the PUF generates an output response of '1'. Conversely, if the clock signal arrives before the D-input signal, the output response is '0'. 
The Arbiter PUF architecture shown in Fig.\ref{FIG:MUX_Arbiter_PUF} generates a 1-bit response for each applied challenge. Multiple Arbiter PUF instances can be combined to increase the output bit length of the overall PUF module.
\subsection{Inverter-based HOA-PUF}
\label{SEC:HOA_PUF}

\begin{figure}[!ht]
	\centering
	\includegraphics[width=0.75\textwidth]{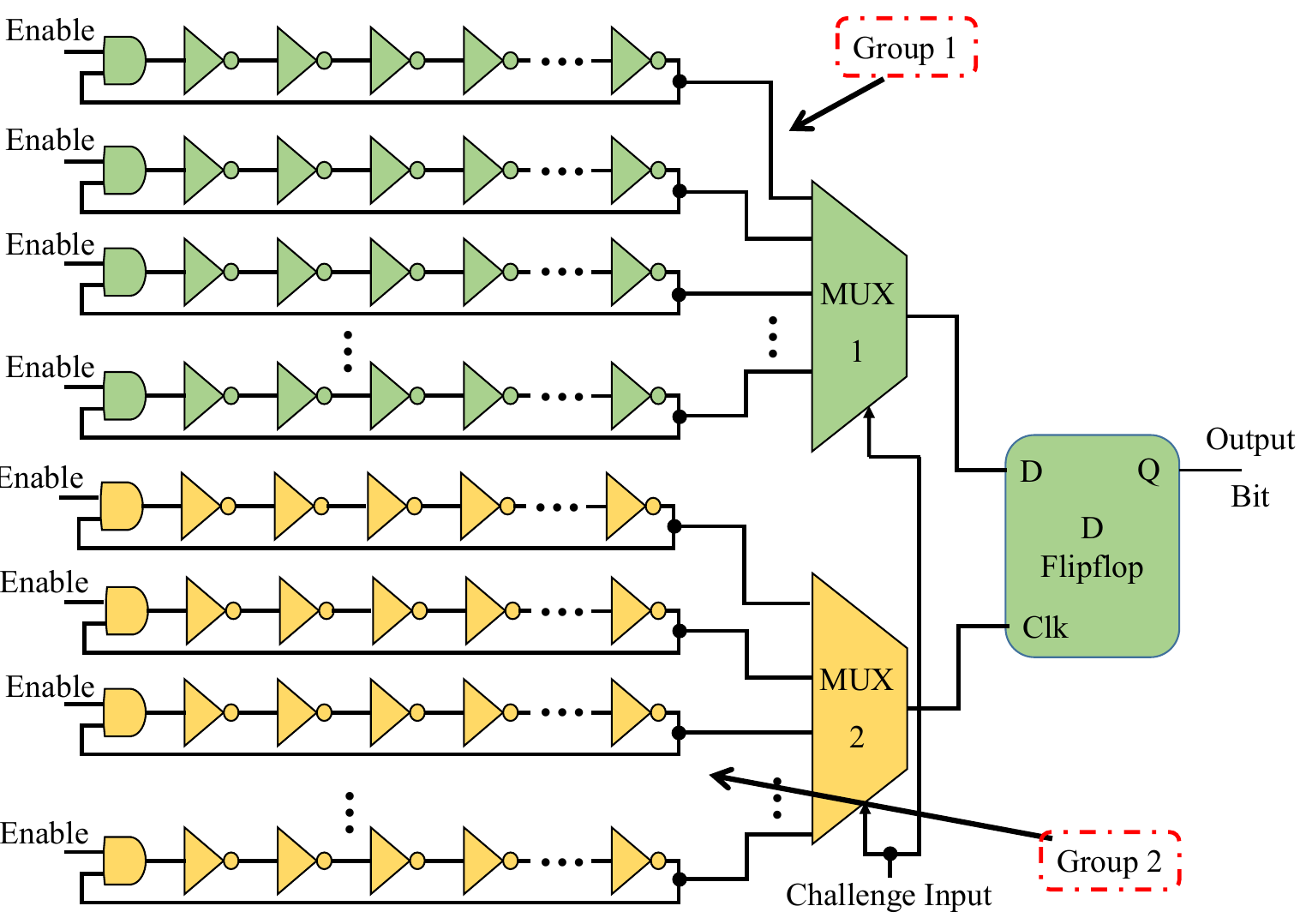}
	\caption{Hybrid Oscillator Arbiter PUF}
	\label{FIG:HOA_PUF}
\end{figure}

Fig. \ref{FIG:HOA_PUF} shows the design of HOA PUF. The core component of the design is the inverter as shown in the figure. Similar to the arbiter PUF, the delay is introduced to the signal by the inverters.  

There are two sets of ROs in the design. All ROs in group 1 are connected to multiplexer 1 (MUX 1), while ROs in group 2 are connected to MUX 2. The architecture and design parameters of the ROs in both groups are identical. However, during the fabrication process, there are unforeseen manufacturing variations in which the output frequencies of individual ROs are not identical. As a result, each RO exhibits a unique oscillation frequency, which contributes to the generation of the PUF response. A D-flip-flop is used as the arbiter in the final stage of the circuit. The output of MUX 1 is connected to the D input of the flip-flop, while the output of MUX 2 is connected to the clock input. The challenge given to the PUF is fed as the select signal to these multiplexers. Unlike Arbiter PUF, this design has multiplexers and requires only one D-Flipflop. 

The challenge decides the ROs that are selected for the output bit. A pair of ROs will generate one output bit similar to the RO PUF. The number of RO pairs selected decides the length of the output key. The length of the key varies with the design and the number of RO stages used in the architecture. Although there can be an exponential number of combinations of ROs, which gives a huge number of keys, not all of them can be used for applications. FoMs must be satisfied before they can be used for cryptographic purposes. 

\begin{figure}[!ht]
		\centering
		\includegraphics[width=0.75\textwidth]{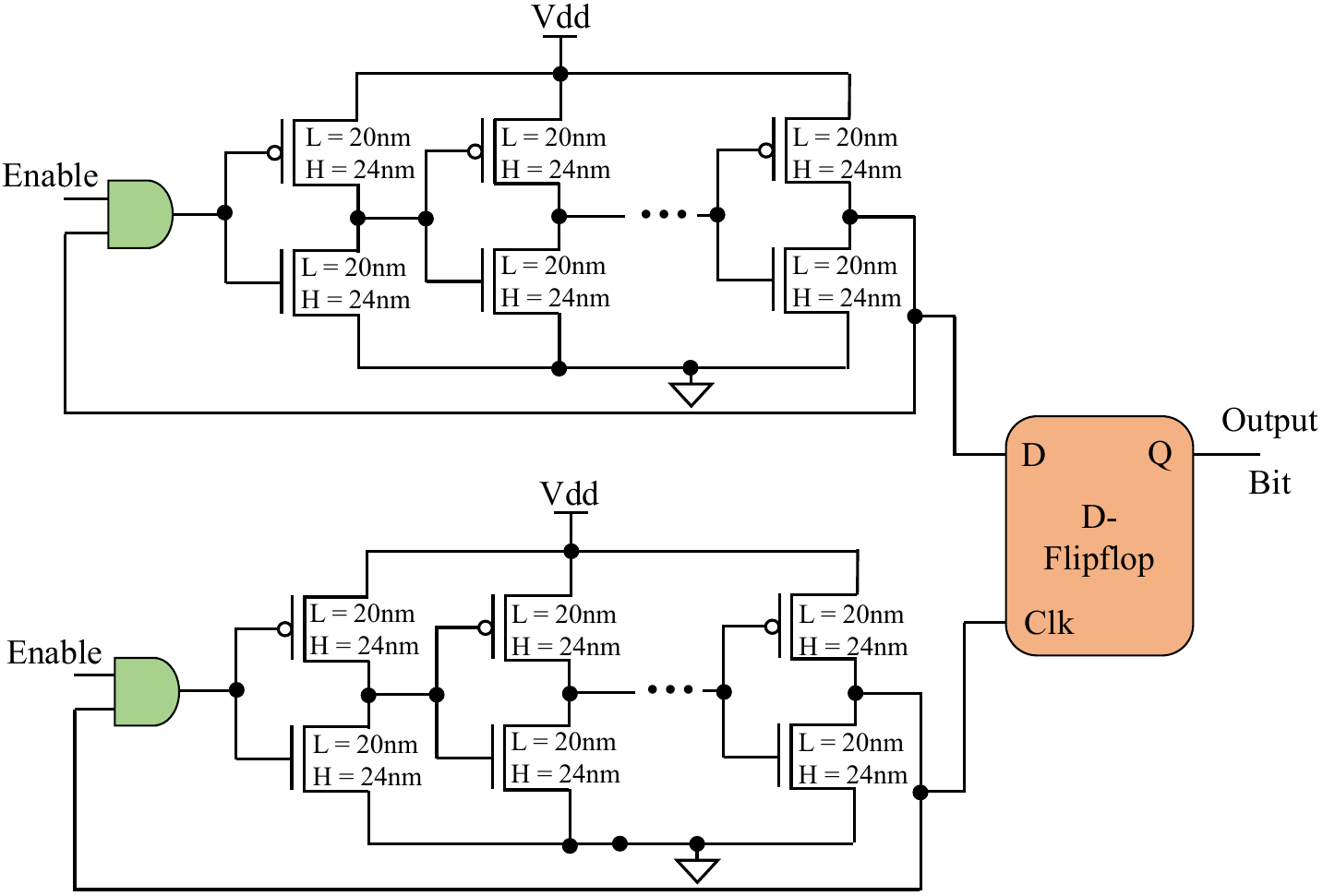}
		\caption{Design of Inverter based HOA PUF.}
		\label{FIG:FinFET_Ring_Oscillator}
\end{figure}

Fig. \ref{FIG:FinFET_Ring_Oscillator} illustrates the FinFET-based 1-bit HOA design. As shown, the inverter forms the core building block of the proposed architecture. The device dimensions, including the channel length and width used in the simulation, are also specified in the figure.

	%\begin{center}
	%	\begin{table}[!h]
	%		\centering
	%		\caption{Nominal values for the FinFET parameters}
	%		\label{Table:nominal_values_finfet}
	%		\begin{tabular}{|c|c|}
	%			\hline
	%			Parameter & Nominal Value\\
	%			\hline\hline
	%			pFET Length & 32n \\
	%			nFET Width & 240n\\
	%			nFET Length & 32n\\
	%			pFET Width &  12n\\
	%			pFET Threshold Voltage & -250mV\\
	%			nFET Threshold Voltage & 310mV\\
	%			pFET Oxide Thickness & 1.65n\\
	%			nFET Oxide Thickness & 1.75n\\
	%			Supply Voltage & 0.9V\\
	%			\hline
	%		\end{tabular}
	%	\end{table}
	%\end{center}

	%Following are the nominal values considered for this design:

\begin{figure}[!ht]
		\centering
		\includegraphics[width=0.75\textwidth]{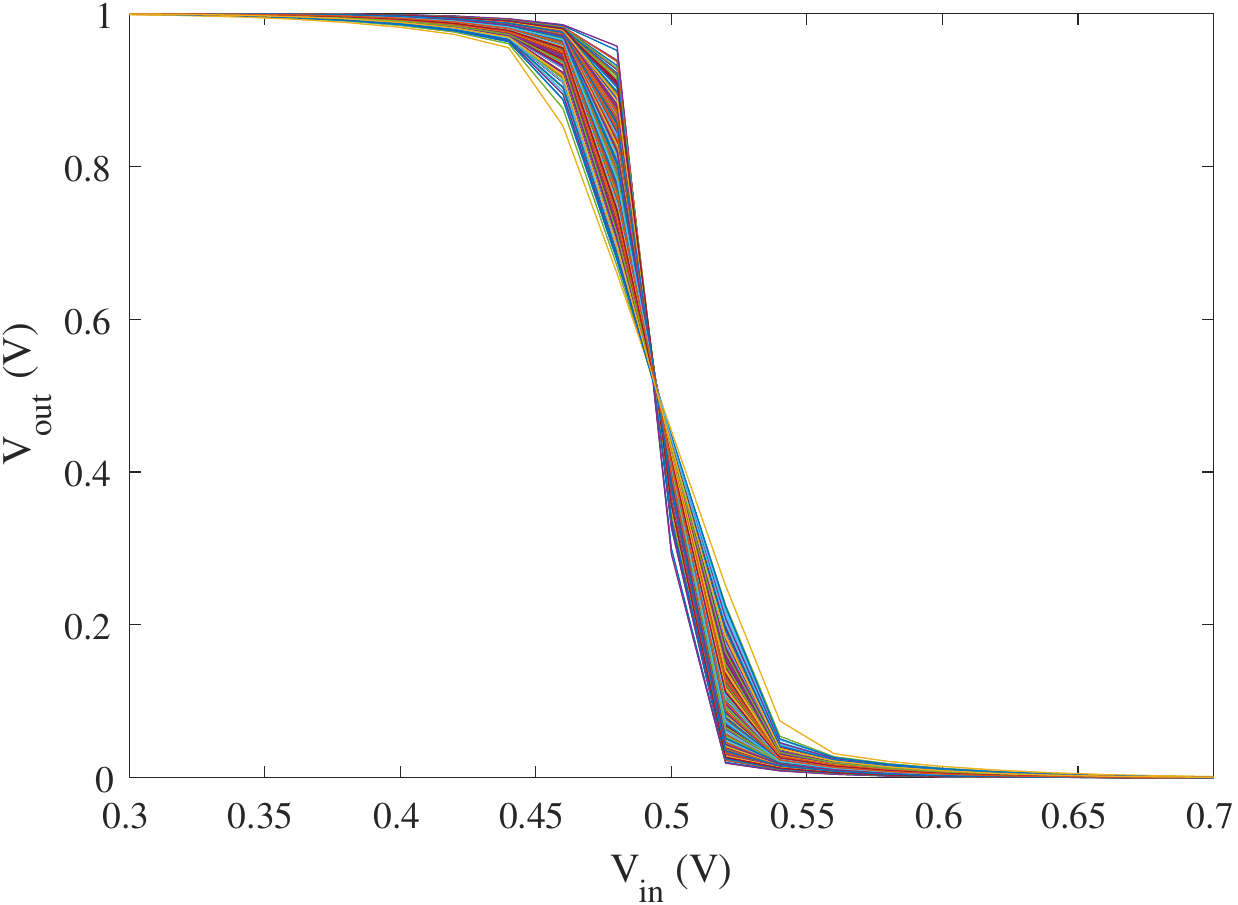}
		\caption{DC Analysis of Inverter with Process and Mismatch Variations.}
		\label{FIG:INV_DC}
\end{figure}

%%%%%%%%%%%%%%%%%%%%%%%%%%%%%%%%%%%%%%%%%%%%%%%%%%%%%%%%%%%%%%%%%%%%%%%%%%%%%%%%
\section{Experimental Results}
\label{SEC:results}

%The experimental results are presented for the two designs of PUF, the Speed Optimized Inverter based MKG PUF and the Power Optimized Inverter based MKG PUF. At the circuit level design of both the PUF, D-Flipflop was implemented. But in this case, an RS-Flipflop could also be implemented. An RS-Flipflop gives less biasing compared to the D-Flipflop. But the RS-Flipflop could not give optimum output key which could generate good results that could satisfy FoMs.

This section presents the simulation results for the designs of Hybrid Oscillator Arbiter PUF and the Arbiter PUF. A comparison of the designs is performed and evaluated for integrating the architecture into the wearable medical designs for a robust security.

\subsection{Experimental Setup}
\label{sec:Simulation_Setup}

The models from \cite{FinfetPDK} were used for simulating the HOA PUF. The FinFET parameters used for the simulation are given in Table \ref{Table:FinFETparams}. The same parameters were taken for all the transistors across all the stages of the ROs. Manufacturing variations were simulated using Monte Carlo simulations. The nominal values presented in Table \ref{Table:FinFETparams} were varied 10 \% to simulate the manufacturing variations. The characteristics of the inverter used in the proposed design are shown in Fig. \ref{FIG:INV_DC}. 

% The simulation of PUFs were performed using the models provided in \cite{FinfetPDK}. In both the designs all the stages of core components, the parameters of the transistors are identical. Monte Carlo simulations were performed on the designs. Monte Carlo simulations are used to vary a parameter in a range provided the mean, deviation and the type of variation. All the geometric properties of the transistors were varied during the simulations. Table \ref{Table:FinFETparams} shows the nominal parameters used for the simulation setup. The nominal parameters were varied with a sigma of 10\% using the Monte Carlo Variation to simulate the manufacturing variations. Fig. \ref{FIG:INV_DC} shows the variations of the inverter characteristics during the simulation of HOA PUF. 

%\begin{center}
	\begin{table}[htbp]
		\centering
		\caption{Nominal Device Parameters Used for the 14 nm FinFETs.}
		\label{Table:FinFETparams}
		\begin{tabular}{|c|c|}
			\hline
	\textbf{Device	Parameters} & \textbf{Specific Values} \\
\hline \hline
			Length (L) & 20 nm \\
			\hline
			Height of Fin & 24 nm \\
			\hline
			Fin Thickness & 10 nm \\
			\hline
			Fin Width & 10 nm\\
			\hline
			Number of Fins & 2\\
			\hline
			Oxide Thickness & 1.3 nm\\
			\hline
			Body Thickness & 1.9 nm\\
			\hline
		\end{tabular}
	\end{table}
%\end{center}

\subsection{Performance Evaluation}

As mentioned in the previous sections, not all the keys generated by the PUF can be used for cryptographic purposes. The following FoMs must be satisfied before they can be used reliably:

% For each design of PUF, 500 runs were simulated and keys were generated. Before the keys can be used for cryptographic applications, some Figures of Merit (FoM) have to be satisfied. Following are the FoMs that have to be analyzed for PUF keys:

	\begin{itemize}
		\item Uniqueness
		\item Reliability
		\item Average Power Consumed
	\end{itemize}

\subsubsection{Uniqueness}
\label{sec:SimResults_Uniqueness}

\emph{Uniqueness} is the PUF's property to generate a unique key for that specific module. It can be categorized as two types of Uniqueness:
\begin{itemize}
	\item \emph{Intra-Uniqueness} - Given a single PUF module $P_1$, every challenge given must produce a unique response.
	\item \emph{Inter-Uniqueness} - When the same challenge input is given to multiple PUF modules, every module is expected to produce a unique response.
\end{itemize}

Uniqueness of a PUF is calculated using hamming distance. When two binary keys are considered, the number of bit positions they differ from each other is the hamming distance between the two keys. The ideal hamming distance between the considered keys must be 50 \%.

% The two properties mentioned above of a unique PUF require analyzing the generated keys from the module. Hamming distance between the generated keys is calculated in each case. Given two binary strings, hamming distance is the number of positions the strings differ from each other. The ideal hamming distance for a key to be used for cryptographic applications is 50 \%. 

\begin{figure}[htbp]
		\centering
		\includegraphics[width=0.46\textwidth]{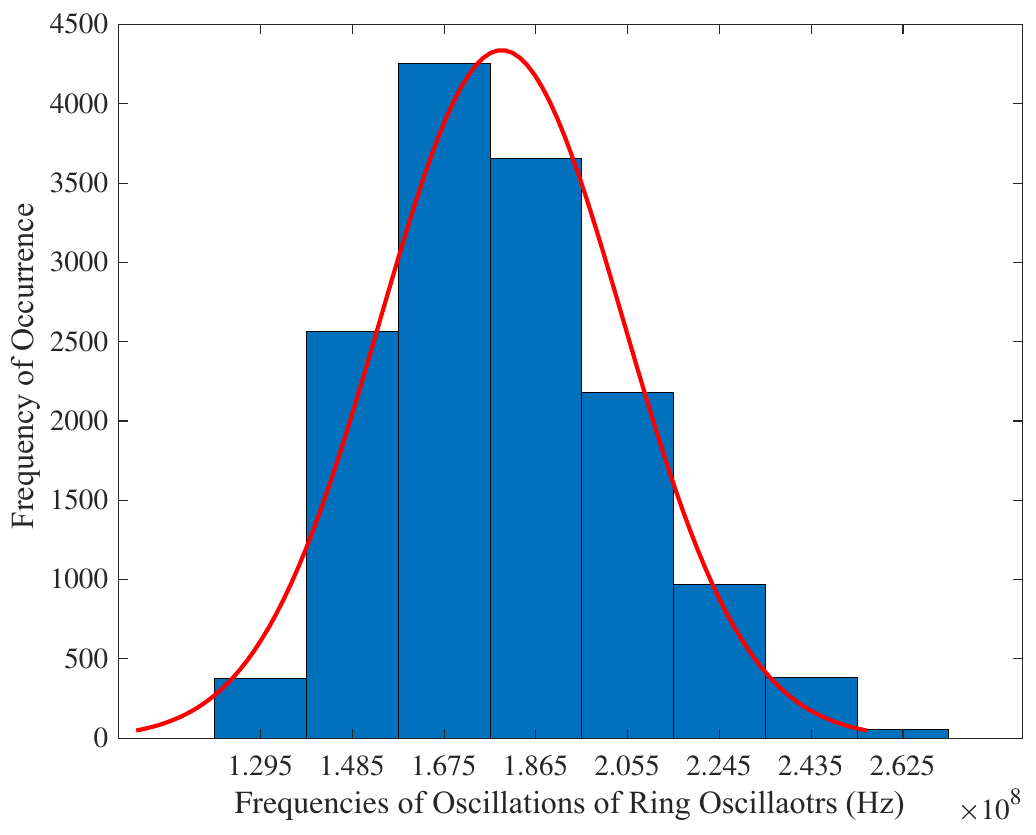}
\caption{Histogram of Oscillation Frequencies of the Ring Oscillators in PUF for 10,000 Oscillators.}
		\label{FIG:ROfrqs}
\end{figure}

\begin{figure}[htbp]
	\centering
	\subfigure[Uniqueness of HOA PUF.]{\label{fig:Uniqueness_HOAPUF}\includegraphics[width=0.45\textwidth]{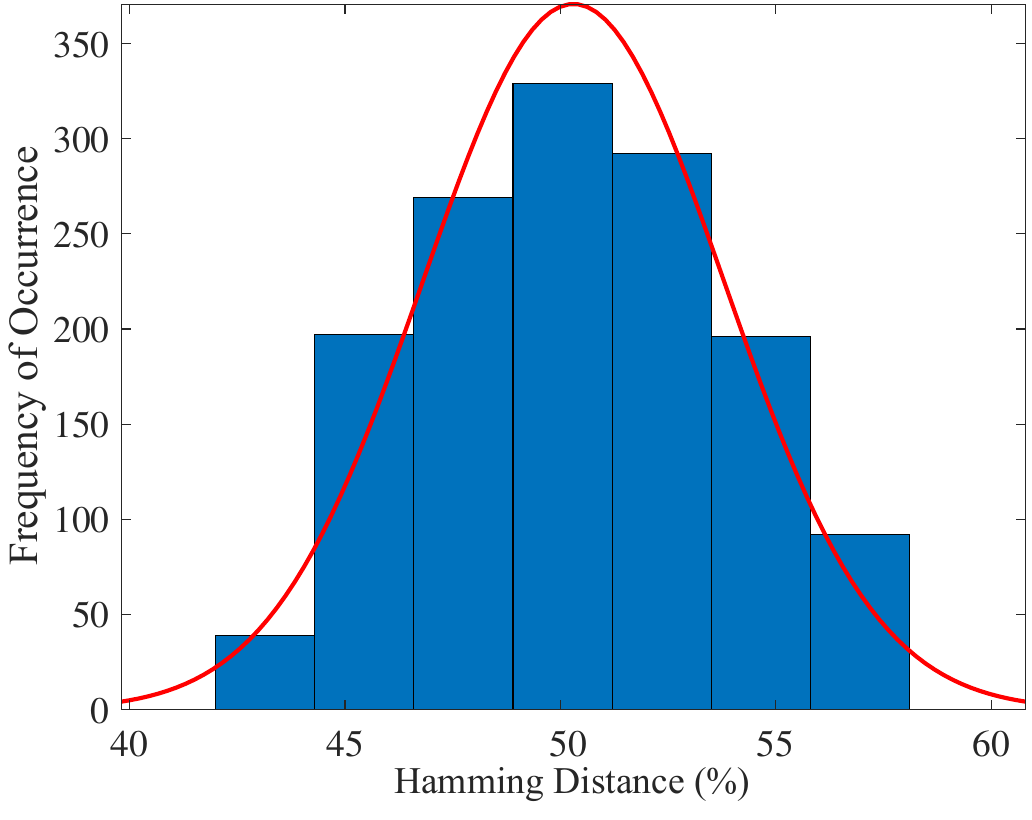}}
	\subfigure[Uniqueness of RO PUF.]{\label{fig:Uniqueness_ROPUF}\includegraphics[width=0.45\textwidth]{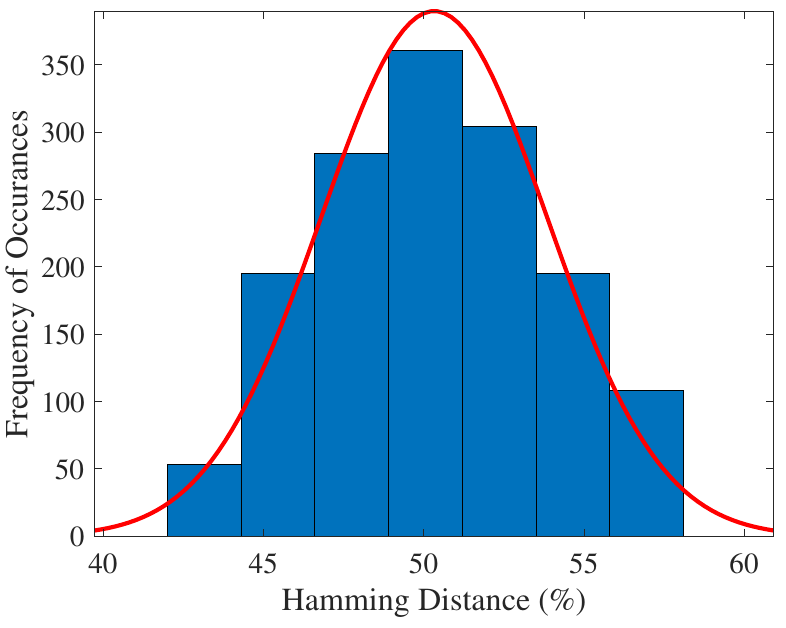}}
	\subfigure[Uniqueness of Arbiter PUF.]{\label{fig:Uniqueness_Arbiter}\includegraphics[width=0.45\textwidth]{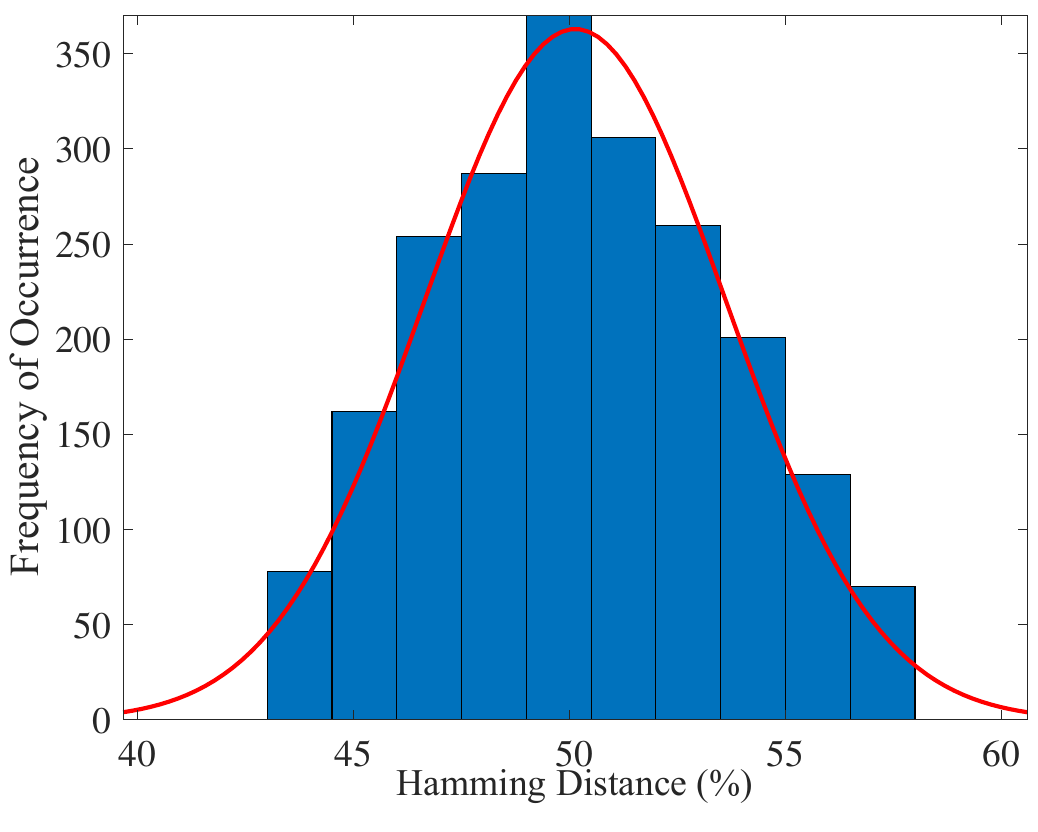}}
	\caption{Uniqueness of the PUFs.}
	\label{FIG:Uniqueness}
\end{figure}

Fig. \ref{FIG:ROfrqs} shows the signal frequencies generated by the ROs in the PUF module. This helps analyze the variation captured by the oscillators in the HOA PUF.
Fig. \ref{FIG:Uniqueness} shows the uniqueness of all three PUF architecutures, the RO PUF, the Arbiter PUF and the HOA PUF. RO PUF and Arbiter PUF uniqueness is taken as a benchmark to evaluate the proposed HOA PUF. The average uniqueness of Arbiter PUF is 50.30 \%, and the average uniqueness of the RO PUF is 50.50 \%. Compared to these, the keys generated by the HOA PUF show a hamming distance of 51.30 \%. All the three values are closer to the ideal value, 50 \%, but the number of keys that are closer to the ideal value are more in the proposed HOA PUF. This shows the HOA PUF successfully captures the manufacturing variations in the circuit.

\begin{table}[htbp]
		\centering
		\caption{Uniqueness of PUF.}
		\label{Table:uniqueness}
		\begin{tabular}{|c|c|c|}
			\hline
PUF Topology & Uniqueness (\%) & Ideal Value (\%)\\
			\hline
			HOA PUF& 51.3 & 50\\
			\hline
			RO PUF & 50.3 & 50\\
			\hline
			Arbiter PUF & 50.3 & 50 \\
			\hline
		\end{tabular}
\end{table}

%	\begin{figure}[!ht]
%		\centering
%		\includegraphics[width=0.45\textwidth]{Images/SpeedInter}
%		\caption{Hamming Distance of Speed Optimized Inverter based MKG PUF.}
%		\label{FIG:SpeedInter}
%	\end{figure}
%	
%	\begin{figure}[!ht]
%		\centering
%		\includegraphics[width=0.45\textwidth]{Images/PowerInter}
%		\caption{Hamming Distance Power Optimized Inverter based MKG PUF.}
%		\label{FIG:PowerInter}
%	\end{figure}
	
%\begin{figure}[!ht]
%		\centering
%		\includegraphics[width=0.45\textwidth]{Uniqueness}
%		\caption{Uniqueness of Ring Oscillator PUF}
%		\label{FIG:Uniqueness_ROPUF}
%\end{figure}
%\begin{figure}[!ht]
%	\centering
%	\includegraphics[width=0.45\textwidth]{Uniqueness_Arbiter}
%	\caption{Uniqueness of Arbiter PUF.}
%	\label{FIG:Uniqueness_Arbiter}
%\end{figure}

\subsubsection{Reliability}
\label{sec:Reliability}

\emph{Reliability} is the ability of a PUF module to generate the same response for a given challenge despite the effects of circuit aging and environmental variations. Once the circuit is fabricated and deployed in a mission-critical application or consumer applications, it is expected to maintain consistent performance throughout the product's operational lifetime. Cryptographic applications, in particular, demand a high degree of reliability and robustness, as the loss or alteration of a cryptographic key may result in data loss and, in some cases, irreversible consequences. Therefore, a PUF is expected to consistently reproduce the same key under varying environmental conditions, including fluctuations in power supply voltage and temperature. 

% \emph{Reliability}  is the property of PUF to generate the same response for a challenge reliably over the time. There are many factors that affect the performance of PUF. The ICs are vulnerable to aging effects. The CRPs change with aging effects. Many aging resistant architectures of PUF are available for IoT architectures. Others which affect the working of the PUF are environmental factors such as power supply and temperature variations. 

%\emph{Reliability} of a PUF is the ability of the module to generate the same key again with the same challenge bits. But the presented circuit is used for generating multiple keys. Hence in this scenario, the reliability of PUF is the ability to generate a new key every time the PUF is run and generate as few interferences as possible. Fig. \ref{fig:SpeedIntra} and Fig. \ref{fig:PowerIntra} present the Hamming distance distribution of the Speed Optimized and Power Optimized Inverter based MKG PUFs. 
	
%To validate the reliability of the circuit, the variables of the transistor are kept constant. The parameters that are varied are only the supply voltage and the temperature. Here temperature is varied as the circuit or system temperature may vary according the the environmental conditions. Monte Carlo simulations were performed on the circuit and the keys generated are tested for uniqueness with the hamming distance. 

\begin{figure}[htbp]
		\centering
		\subfigure[Reliability of HOA PUF.]{\label{fig:Reliability_HOA}\includegraphics[width=0.45\textwidth]{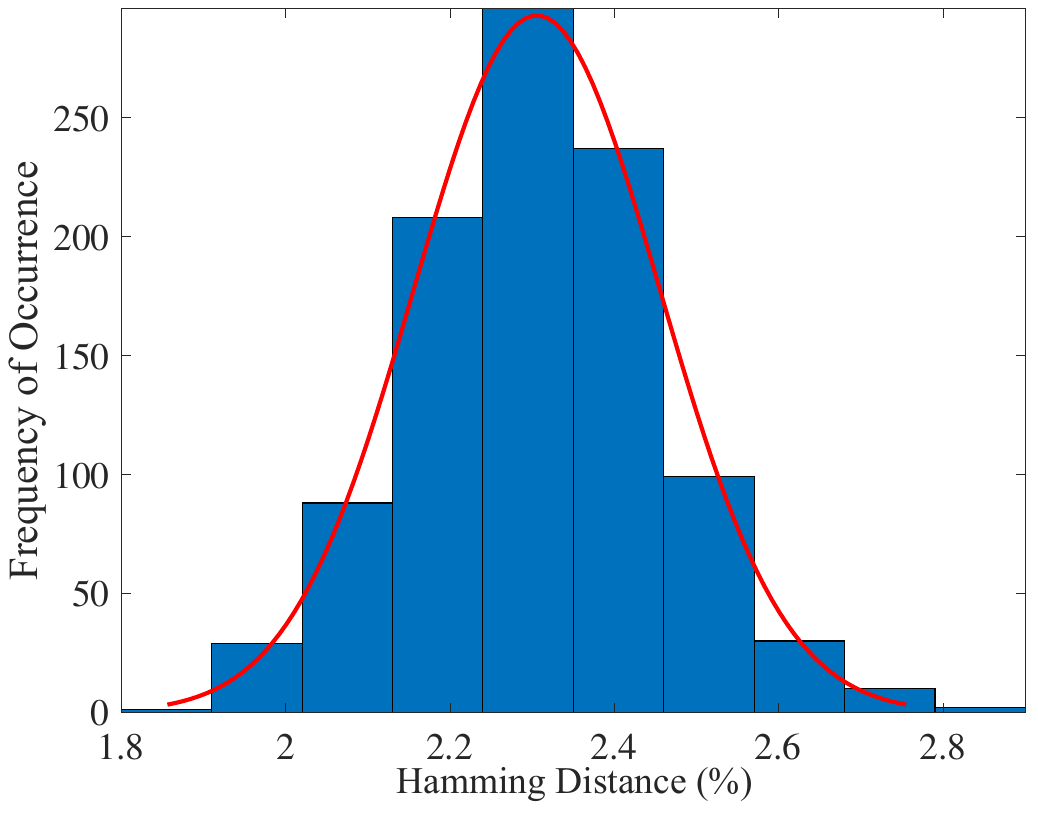}}
		\subfigure[Reliability of RO PUF.]{\label{fig:Reliability_RO}\includegraphics[width=0.45\textwidth]{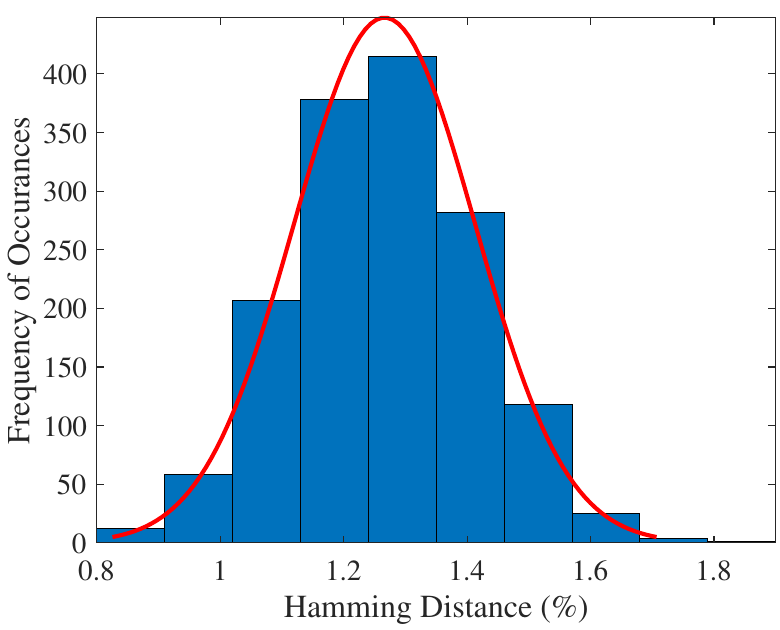}}
		\subfigure[Reliability of Arbiter PUF.]{\label{fig:Reliability_Arbiter}\includegraphics[width=0.45\textwidth]{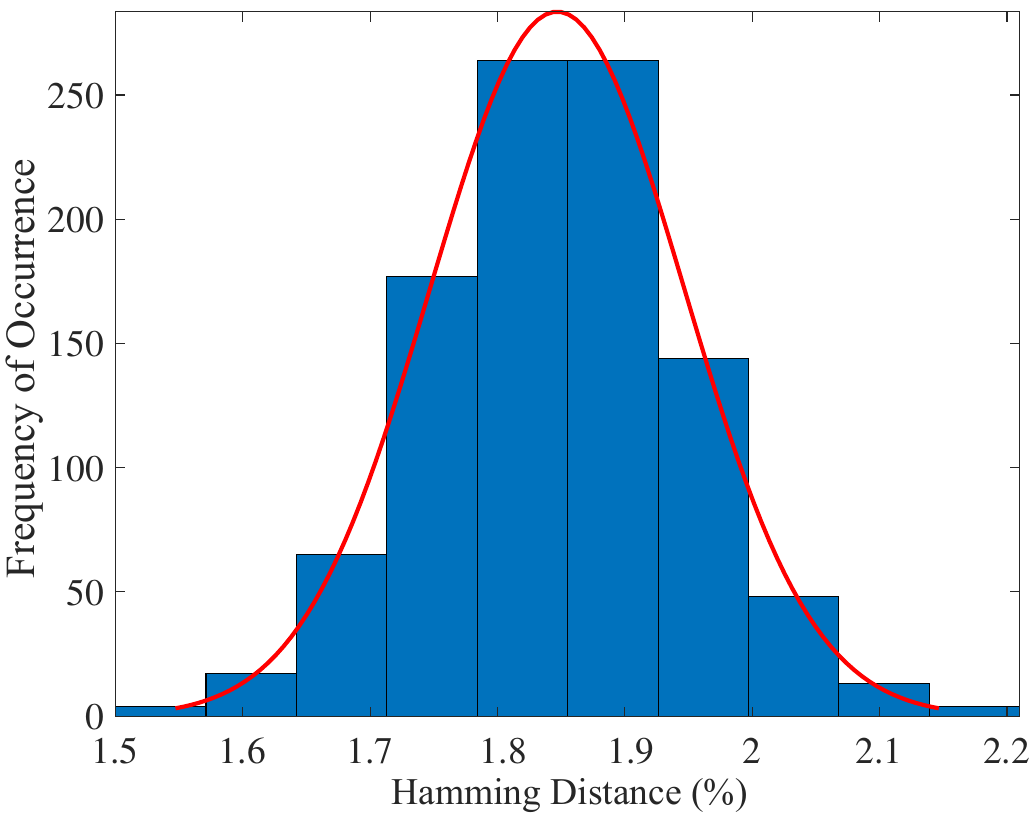}}
		\caption{Reliability of PUFs.}
		\label{FIG:Reliability}
\end{figure}

Fig. \ref{FIG:Reliability} shows the simulation results of the reliability of the PUF architectures. The reliability of the PUF is also measured using hamming distance. During simulation, the same challenge is given to the PUF  under temperature and power supply variations. As shown in the figure, the error rate of the HOA PUF is 2.35 \%, RO PUF is 1.32 \%, and Arbiter PUF is 1.8 \% which are closer to the ideal values of 0 \%.

% Fig. \ref{FIG:Reliability} shows the simulation results of reliability of the PUF architectures. Reliability of PUFs is also estimated using the hamming distance. For simulation, same challenge input is given to a PUF module by varying other parameters such as temperature and power supply. As shown in the figure, the error rate of the PUF modules is 2.35 \% for HOA PUF, 1.32\% for RO PUF, and 1.8 \% for Arbiter PUF. 

% \textbf{\textcolor{red}{Rewrite Below}}\textcolor{red}{Rewrite Below}

\subsubsection{Average Power}
\label{sec:Average_Power}

Wearable devices require ultra low power operation for a longer battery life. The power consumption of a PUF module plays a crucial role in its integration into applications in wearable technology. The average power consumed is the average of the total power consumed by the module over the period of key generation. In addition to wearable devices, the power consumption also plays an major role in the integration of this PUF design into resource-constrained environments, including smart healthcare and smart city applications.  

% The average power consumed is the average of the total power consumed by the module over the period of key generation. This is one of the key aspects of security modules for IoT applications. As discussed in previous sections, the IoT devices are deployed in places where it is not constantly monitored by users. Some devices will also have to run on battery. Hence a low power overhead is required in security. 

\begin{figure}[htbp]
		\centering
		\subfigure[Average Power Consumption of HOA PUF.]{\label{fig:AveragePower_HOA}\includegraphics[width=0.40\textwidth]{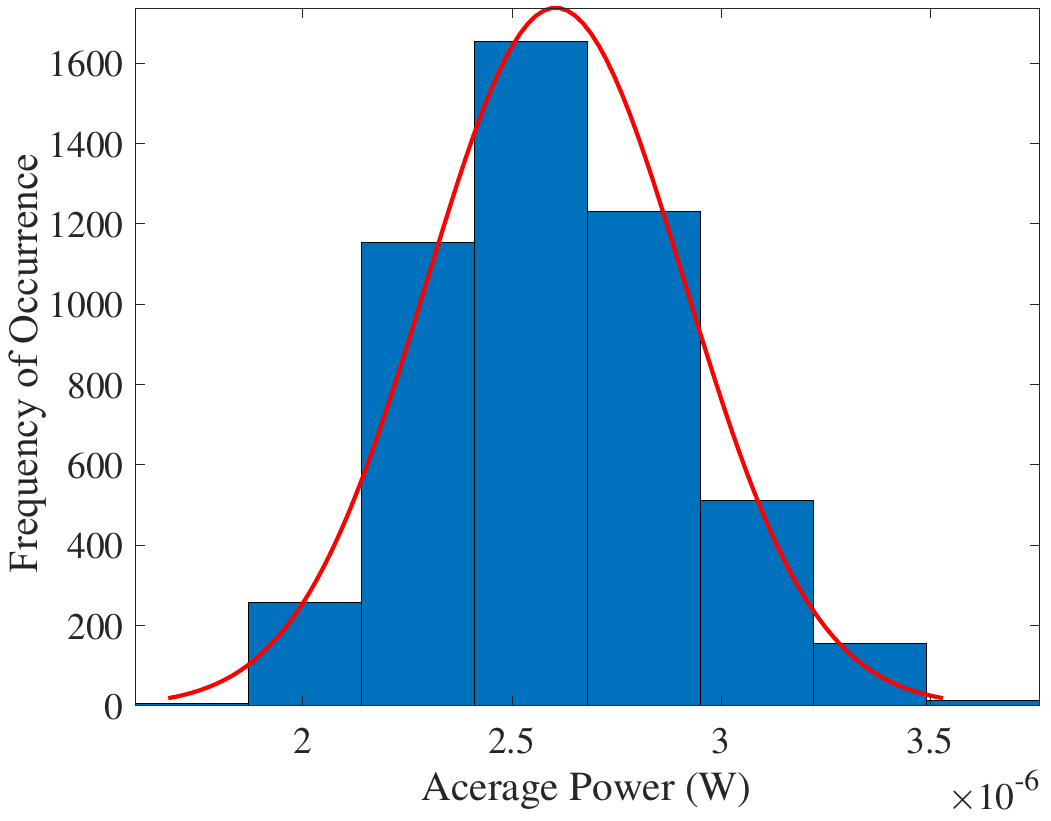}}
		\subfigure[Average Power Consumption of RO PUF.]{\label{fig:AveragePower_RO}\includegraphics[width=0.4\textwidth]{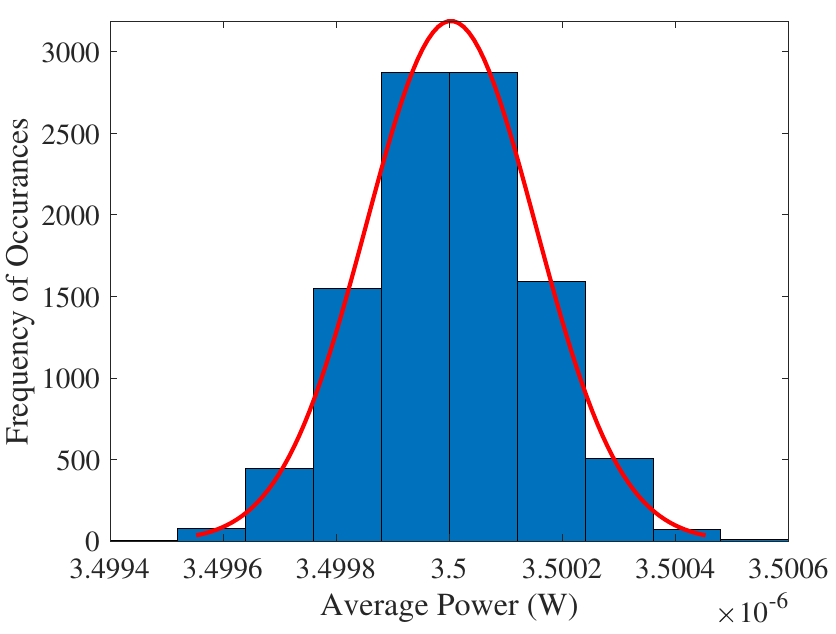}}
		\subfigure[Average Power Consumption of Arbiter PUF.]{\label{fig:AveragePower_Arbiter}\includegraphics[width=0.4\textwidth]{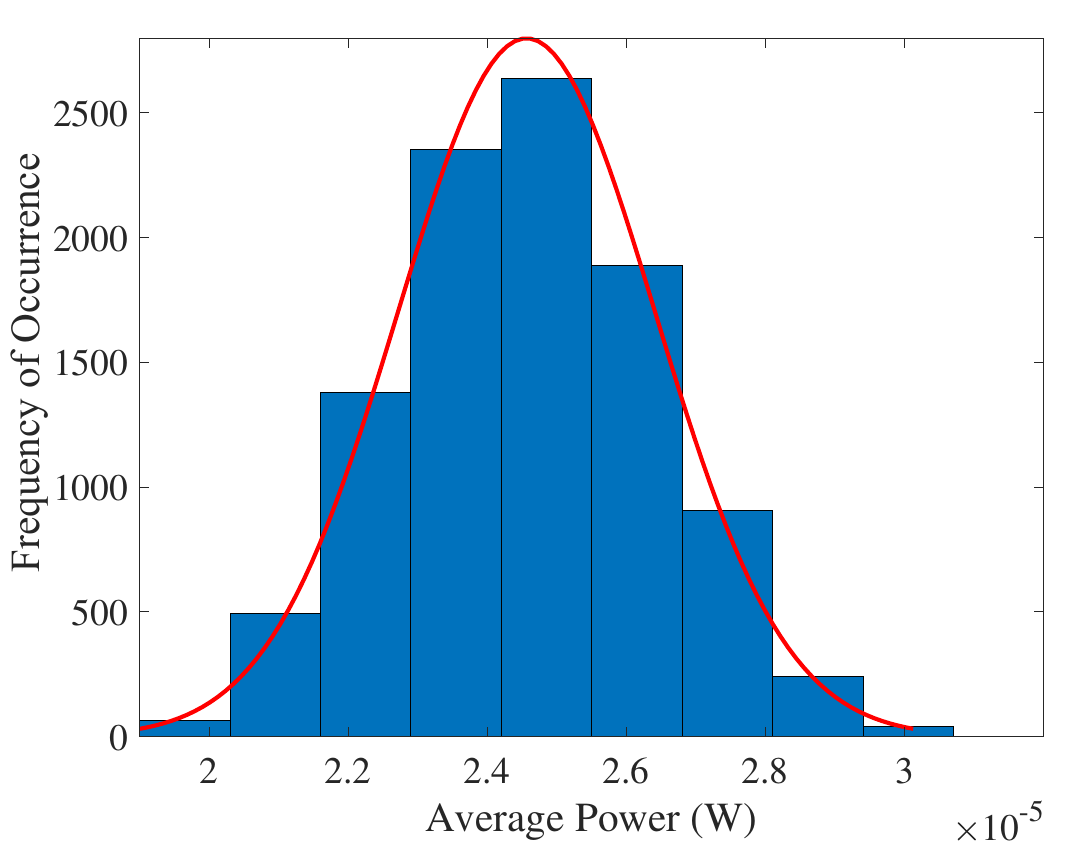}}
		\caption{Average Power Consumption of the Proposed PUFs.}
		\label{FIG:AveragePower}
\end{figure}

The average power consumption of all three PUF architectures is shown in Fig. \ref{FIG:AveragePower}. The Arbiter PUF has 16 stages of Multiplexers and HOA PUF has 7 stages of inverters. The RO PUF and the HOA PUF have the same number of inverters with the only difference being an additional counter and comparator. 
The benchmark results were generated for both the RO PUF and the Arbiter PUF. RO PUF consumes about 3.5 $\mu$W and the arbiter PUF consumes about 25 $\mu$W. The HOA PUF on the other hand, consumes about 2.7 $\mu$W to generate 1 key. 

% Fig. \ref{FIG:AveragePower} shows the power consumption of PUF architectures. The simulations were run and the power consumption for both the architectures was measured when the keys are being generated. As shown in the figure, the total power consumption is around 2.7 $\mu$W for HOA PUF, 3.5 $\mu$W for RO PUF and 2.5 $\mu$W for Arbiter PUF. The Arbiter PUF has 16 stages of Multiplexers and HOA PUF has 7 stages of inverters. The RO PUF and the HOA PUF have the same number of inverters with the only difference of an additional counter and comparator. 

\subsection{Analysis of PUF Architectures}
\label{sec:Analysis}

Table \ref{Table:Characterization_of_PUF} shows the characterization of the proposed HOA PUF and a comparison with the other two architectures, RO PUF and Arbiter PUF. The uniqueness of the HOA PUF is about 51.3 \% compared to the RO PUF, which is 50.4 \% and the arbiter PUF, which is 50.3 \%, as shown in the table. Although the average uniqueness of the RO PUF and the Arbiter PUF are closer to the ideal value of 50 \%, as mentioned before, the standard deviation is higher in keys generated by the other two architectures. In addition to the lower standard deviation of average uniqueness, the error rate of the HOA PUF is 2.35 \%. The average power consumption of the HOA PUF is also significantly lower compared to the other two architectures, and Arbiter PUF in particular which is about 10 $\times$ higher than the HOA PUF. As the Arbiter PUF does not contain a multiplexer dedicated to selection of the individual components, the key generation time is much higher in the arbiter PUF. But there is also a trade off of the power consumption in the Arbiter PUF. 

% The characterization table of the architectures is shown in Table \ref{Table:Characterization_of_PUF}. A consolidated table of simulation results is shown here. The uniqueness of HOA PUF compared to the Arbiter PUF is better in the results. But the number of keys generated around the mean value is more in case of the Arbiter PUF. Reliability of Arbiter PUF is high compared to the HOA PUF. The error rate of RO PUF is 1.32 \%, Arbiter PUF is 1.8 \% unlike the error rate of HOA PUF which is 2.35 \%. The power consumption of Arbiter PUF is 10 times that of the HOA PUF and RO PUF due to the presence of more number of devices. When the time necessary to generate the keys is considered, the Arbiter PUF can generate the keys faster than the HOA PUF as there involves a Multiplexer selecting the oscillators. High performance preferred applications that can trade-off power can utilize Arbiter PUF but in low power devices, trading off the performance, HOA PUF or RO PUF can be integrated.

\begin{table}[htbp]
			\centering
			\caption{Characterization Of Power Optimized PUF Designs.}
			\label{Table:Characterization_of_PUF}
			\begin{tabular}{|c|c|c|c|c|}
				\hline
%				\multicolumn{5}{|c|}{Power Optimized Inverter MKG PUF} \\
%				\hline
\textbf{Metrics} & \textbf{Ideal Value} & \textbf{HOA PUF} & \textbf{RO PUF}& \textbf{Arbiter PUF} \\
				\hline
\hline
				Uniqueness & 50 \% & 51.3 \% & 50.4 \% & 50.3 \% \\
				\hline
				Reliability & 0 \% & 2.35 \% & 1.32 \% & 1.8 \% \\
				\hline
				Average Power & -- & 2.7 $\mu$W & 3.5 $\mu$W & 25 $\mu$W\\
				\hline
			\end{tabular}
\end{table}

\begin{table*}[htbp]
		\centering
		\caption{Comparison of Results with Related Existing Research.}
		\label{Table:publication_comparison}
		\begin{tabular}{|p{4cm}|p{3cm}|p{3.3cm}|p{2cm}|p{1.6cm}|}
			%		\hline
			%		Parameter & \multicolumn{3}{|c|}{Value}\\
			\hline
			%			& \multirow{2}{*}{Traditional RO PUF} & \multirow{2}{*}{\parbox{3cm}{Speed Optimized Hybrid Oscillator Arbiter PUF}}&\multirow{2}{*}{\parbox{3cm}{Power Optimized Hybrid Oscillator Arbiter PUF}} \\ 
			%			Transistor sizes &&& \\\cline{2-4}
			%			&	120n : 32n & 240n : 32n&\\
\textbf{Research Works} & \textbf{Technology Used} & \textbf{Architecture Used} & \textbf{Average Power Requirement} & \textbf{Hamming Distance (\%)}\\
			\hline\hline
			Rahman et al. \cite{Aging_ROPUF} & 90 nm& & -- & 50\\
			\hline
			Maiti et al. \cite{Improved_PUF_Journal} & 180 nm & Traditional Ring Oscillator & -- & 50.72 \\
			\hline
			Garrett et al. \cite{Garrett_DAC_2015} & Memristor & Memristor Crossbar & 564.83 $\mu$W & 49.96\\
			\hline
			Uddin et al. \cite{Uddin_ISVLSI_2016} & Memristor & XORed Memristor Crossbar & 450 $\mu$W & 50\\
			\hline
			Suh et al. \cite{PUF_ACM_2007} & -- &  & -- & 46.15 \\
			\hline
			Maiti et al. \cite{puf_characterization} & -- & -- & -- & 47.31\\
			\hline
			Mathew et al. \cite{Mathew_ISSCC_2014} & 22 nm CMOS & Hybrid Delay Cross Coupled & -- & 50.14 \\
			\hline
			Suh et al. \cite{Suh_DAC_2007} & 90 nm & Ring Oscillator & -- & 50 \\
			\hline
			Cao et al. \cite{Cao_TCS_2015} & 180 nm & CMOS Image Sensor PUF & -- & 49.37\\
			\hline
			
			Yanambaka et al. \cite{inis_prasanth_2016} & 32 nm FinFET & Inverter based MKG PUF& 251 $\mu$W & 48.3\\
			\hline
			Yanambaka et al. \cite{inis_prasanth_2016} & 32 nm FinFET& Current Starved Oscillator & 320 $\mu$W & 50.9\\
			\hline
			Nozaki et al. \cite{Nozaki_GCCE_2019} & 180 nm & Ring Oscillator PUF & -- & 63.83 \\
			\hline
			Sahoo et al. \cite{Sahoo_IEEETC_2018} & 65 nm & Arbiter PUF &  -- & 50.04\\
			\hline
            Zayed et al. \cite{Zayed_IEEE_Access_2021} & 20nm FinFET & Oscillator Collapse & 140 nW - 740 nW & 50.01 \% \\
            \hline
            Driemeyer et al. \cite{Driemeyer_IEEEJSSC_2026} & 28-nm CMOS & Hybrid RO/Arbiter PUF & 0.259 pJ/bit & 51.80 \% \\
            \hline
            Pahlevi et al. \cite{Pahlevi_IEEEAccess_2025} & - & Pre-Selection–Enhanced Arbiter PUF & - & 10 \% - 50 \% \\
            \hline

            Das et al. \cite{Das_IEEEJN_2025} & 45 nm & voltage-gated spin-orbit torque magnetic tunnel junctions (VGSOT-MTJs) based PUF & 63.67 fJ/bit & 50.20 \%  \\
            \hline
			\textbf{This Paper (Arbiter PUF)} & 14 nm FinFET & Multiplexer & 25 $\mu$W & 50.3\\
			\hline
			\textbf{This paper (HOA PUF)} & 14 nm FinFET & Traditional Ring Oscillator & 2.7 $\mu$W & 51.3\\
			\hline
			
		\end{tabular}
\end{table*}

\section{Conclusion and Future Research}
\label{SEC:conclusion}

This paper presented the analysis of two designs of PUF, Arbiter PUF and Hybrid Oscillator Arbiter PUF for integration into wearable devices. These wearable devices are low power devices that are incapable of performing high cryptographic tasks. Introducing such a PUF module to these devices helps generate keys and eliminates the need to store the keys. Depending on the PUF module architecture, it can generate any length of keys, and, if necessary, the PUF module is reconfigurable in some cases. The power consumption of Arbiter PUF is 25 $\mu$W and HOA PUF is 2.7 $\mu$W.  The uniqueness of the PUF modules is also 51.3 \% and 50.3 \% which are near the ideal values. These two designs are ideal to be implemented in an appropriate IoT ecosystem to increase security.
The future directions of this research include implementing PUF designs that are capable of more reconfigurability and ultra-low power design. Moreover, these designs will be integrated into an IoT ecosystem and a prototype will be designed for further analysis.

%\textbf{\textcolor{red}{
%Future research ... \\
%Multi-key PUF ... \\
%Re-configurable PUF ... \\
%}}
%
%The future directions of this research is in multiple fronts. We intend to explore the quality of PUF design when it is realized using junction and doping free transistors \cite{Panchore_iNIS-2016_JLFET}. We want to investigate alternative topologies for PUF circuit design which can be robust and energy-efficient. It will be interesting to evaluate the performance of specific PUFs for different nanoelectronic technology including FinFET, GNR-FET, and other similar technologies. The deployment of PUF in real-life systems to provide efficient security for smart healthcare and even in smart cities in a larger context needs research and development \cite{Prabha_Thyroid}. 

%%%%%%%%%%%%%%%%%%%%%%%%%%%%%%%%%%%%%%%%%%%%%%%%%%%%%%%%%%%%%%%%%%%%
\section{Acknowledgments}
	
The preliminary research has been presented with the 32nm FinFETs in the following peer-reviewed conference \cite{Yanambaka_iNIS_2016_Multi-Key-PUF} and has been changed to the 14nm FinFETs in the current paper.
%Bibliography
\bibliographystyle{unsrt}  
\bibliography{Bibliogrpahy_Delay-Based-PUFs}

% \newpage
\section*{Authors' Biographies}
\noindent
\begin{wrapfigure}{l}{0.22\textwidth}
    \vspace{-6pt}
    \includegraphics[width=0.22\textwidth]{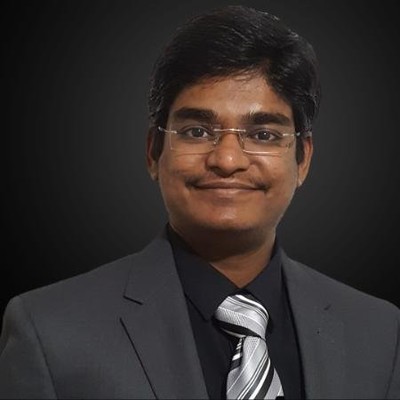}
    \vspace{-10pt}
\end{wrapfigure}
\noindent \textbf{Venkata P. Yanambaka} received the Bachelor of Technology degree in Computer Science and Engineering from Priyadarshini College of Engineering and Technology, Nellore, India, the Master of Science degree in Computer Engineering from the University of North Texas, Denton, TX, and the Ph.D.\ degree in Computer Science and Engineering from the University of North Texas, Denton, TX, in 2019, under the supervision of Dr.\ Saraju P.\ Mohanty. His dissertation, titled ``Exploring Physical Unclonable Functions for Efficient Hardware Assisted Security in IoT,'' was awarded the Best Poster Award at IEEE MetroCon 2017. He is currently an Assistant Professor of Computer Science at Texas Woman's University, Denton, TX. His research interests are in the Internet of Things (IoT), hardware-assisted security primitives, Physical Unclonable Functions (PUFs), and security in IoT architectures. He has authored over 50 peer-reviewed publications, including multiple journal and transaction articles and a book chapter. He has served as a reviewer for several peer-reviewed journals and conferences. He has been an active member of IEEE since 2012.

\begin{wrapfigure}{l}{0.22\textwidth}
    \vspace{-6pt}
    \includegraphics[width=0.22\textwidth]{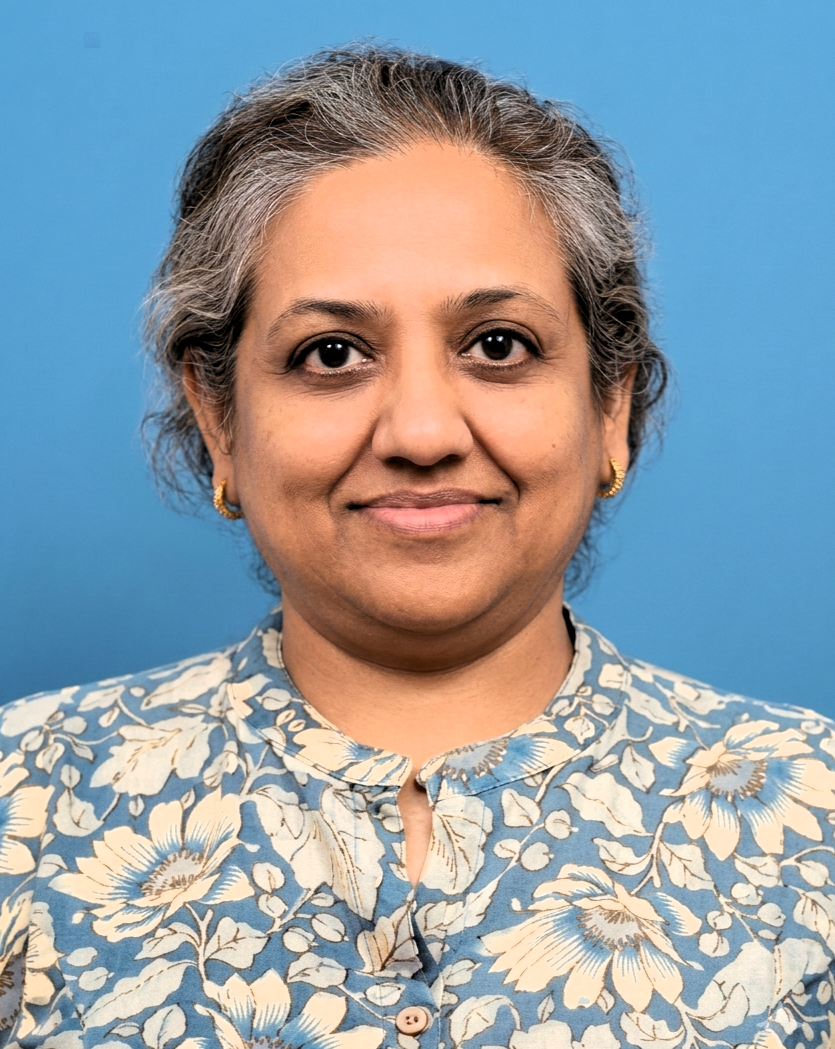}
    \vspace{-10pt}
\end{wrapfigure}
\textbf{Uma Choppali} Uma Choppali received her bachelor’s degree in science and education from the Regional Institute of Education, Bhubaneswar, India, in 1999, the master’s degree in physics from the University of South Florida, USA, in 2004, and the Ph.D. degree in materials science and engineering from the University of North Texas, USA, in 2006. She is a Faculty Member with Dallas College (Eastfield Campus), Mesquite, TX, USA. She has authored a dozen research articles. Her Google Scholar H-index is 12 and i10-index is 13 with 2275 citations. She is a regular reviewer of several international journals and conferences.
\\\\\\\\\\\\

\noindent 
\begin{wrapfigure}{l}{0.22\textwidth}
    \vspace{-6pt}
    \includegraphics[width=0.22\textwidth]{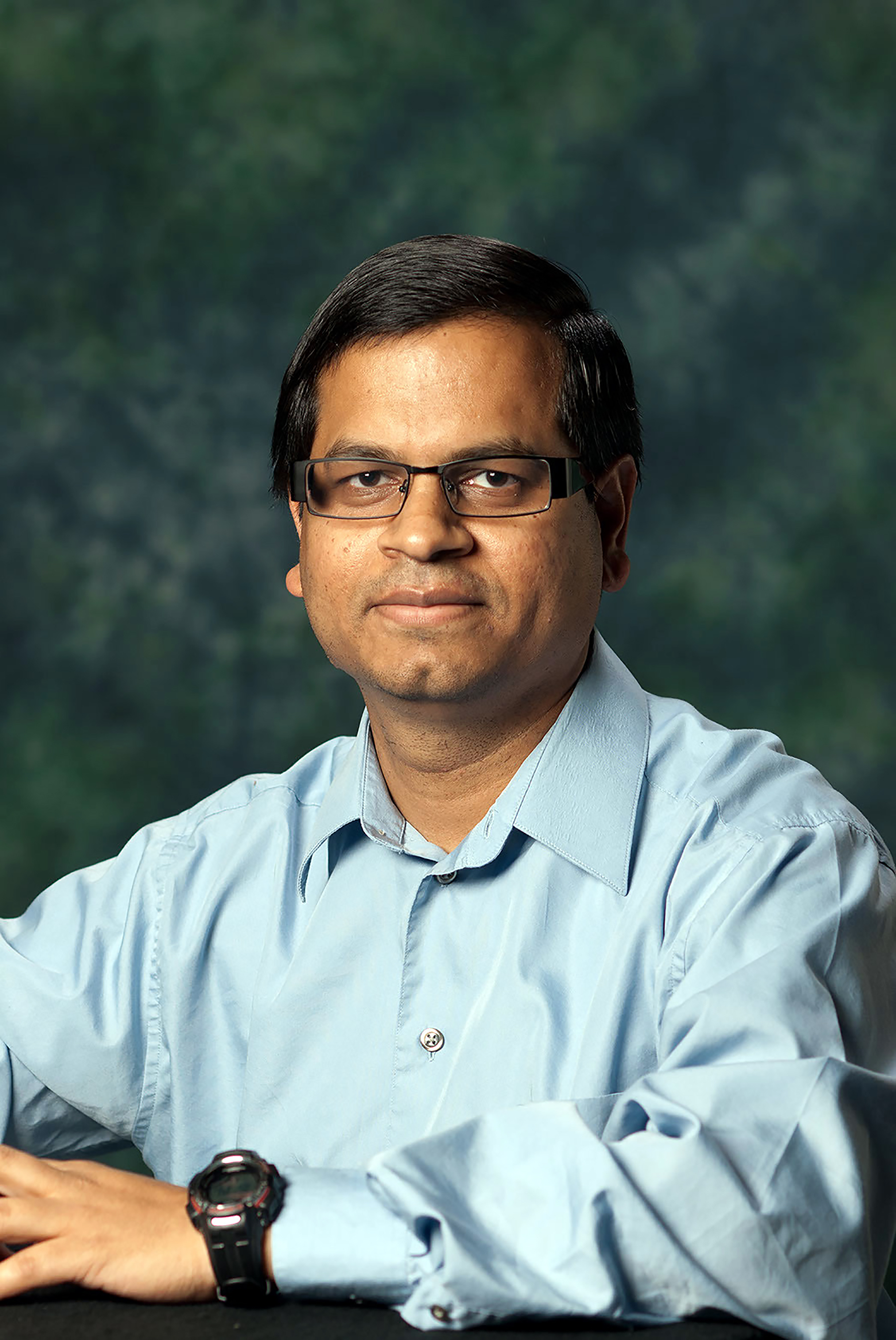}
    \vspace{-10pt}
\end{wrapfigure}
\noindent \textbf{Saraju Mohanty} received the bachelor’s degree (Honors) in electrical engineering from the Orissa University of Agriculture and Technology, Bhubaneswar, in 1995, the master’s degree in Systems Science and Automation from the Indian Institute of Science, Bengaluru, in 1999, and the Ph.D. degree in Computer Science and Engineering from the University of South Florida, Tampa, in 2003. He is a Professor with the University of North Texas. His research is in ``Smart Electronic Systems’’ which has been funded by National Science Foundations (NSF), Semiconductor Research Corporation (SRC), U.S. Air Force, NIDILRR, IUSSTF, and Mission Innovation. He has authored 600 research articles, 5 books, and 10 granted and pending patents. His Google Scholar h-index is 68 and i10-index is 338 with 19,000 citations. He is regarded as a visionary researcher on Smart Cities technology in which his research deals with security and energy aware, and AI/ML-integrated smart components. He introduced the Secure Digital Camera (SDC) in 2004 with built-in security features designed using Hardware Assisted Security (HAS) or Security by Design (SbD) principle. He is widely credited as the designer for the first digital watermarking chip in 2004 and first the low-power digital watermarking chip in 2006. He is a recipient of 21 best paper awards, Fulbright Specialist Award in 2021, IEEE Consumer Electronics Society Outstanding Service Award in 2020, the IEEE-CS-TCVLSI Distinguished Leadership Award in 2018, and the PROSE Award for Best Textbook in Physical Sciences and Mathematics category in 2016. He has delivered 33 keynotes and served on 15 panels at various International Conferences. He has been serving on the editorial board of several peer-reviewed international transactions/journals, including IEEE Transactions on Big Data (TBD), IEEE Transactions on Computer-Aided Design of Integrated Circuits and Systems (TCAD), IEEE Transactions on Consumer Electronics (TCE), and ACM Journal on Emerging Technologies in Computing Systems (JETC). He has been the Editor-in-Chief (EiC) of the IEEE Consumer Electronics Magazine (MCE) during 2016-2021. He served as the Chair of Technical Committee on Very Large Scale Integration (TCVLSI), IEEE Computer Society (IEEE-CS) during 2014-2018 and on the Board of Governors of the IEEE Consumer Electronics Society during 2019-2021. He serves on the steering, organizing, and program committees of several international conferences. He is the steering committee chair/vice-chair for the IEEE International Symposium on Smart Electronic Systems (IEEE-iSES), the IEEE-CS Symposium on VLSI (ISVLSI), and the OITS International Conference on Information Technology (OCIT). He has supervised 3 post-doctoral researchers, 21 Ph.D. dissertations, 29 M.S. theses, and 41 undergraduate projects.  received the bachelor’s degree (Honors) in electrical engineering from the Orissa University of Agriculture and Technology, Bhubaneswar, in 1995, the master’s degree in Systems Science and Automation from the Indian Institute of Science, Bengaluru, in 1999, and the Ph.D. degree in Computer Science and Engineering from the University of South Florida, Tampa, in 2003. He is a Professor with the University of North Texas. His research is in ``Smart Electronic Systems’’ which has been funded by National Science Foundations (NSF), Semiconductor Research Corporation (SRC), U.S. Air Force, NIDILRR, IUSSTF, and Mission Innovation. He has authored 600 research articles, 5 books, and 10 granted and pending patents. His Google Scholar h-index is 68 and i10-index is 338 with 19,000 citations. He is regarded as a visionary researcher on Smart Cities technology in which his research deals with security and energy aware, and AI/ML-integrated smart components. He introduced the Secure Digital Camera (SDC) in 2004 with built-in security features designed using Hardware Assisted Security (HAS) or Security by Design (SbD) principle. He is widely credited as the designer for the first digital watermarking chip in 2004 and first the low-power digital watermarking chip in 2006. He is a recipient of 21 best paper awards, Fulbright Specialist Award in 2021, IEEE Consumer Electronics Society Outstanding Service Award in 2020, the IEEE-CS-TCVLSI Distinguished Leadership Award in 2018, and the PROSE Award for Best Textbook in Physical Sciences and Mathematics category in 2016. He has delivered 33 keynotes and served on 15 panels at various International Conferences. He has been serving on the editorial board of several peer-reviewed international transactions/journals, including IEEE Transactions on Big Data (TBD), IEEE Transactions on Computer-Aided Design of Integrated Circuits and Systems (TCAD), IEEE Transactions on Consumer Electronics (TCE), and ACM Journal on Emerging Technologies in Computing Systems (JETC). He has been the Editor-in-Chief (EiC) of the IEEE Consumer Electronics Magazine (MCE) during 2016-2021. He served as the Chair of Technical Committee on Very Large Scale Integration (TCVLSI), IEEE Computer Society (IEEE-CS) during 2014-2018 and on the Board of Governors of the IEEE Consumer Electronics Society during 2019-2021. He serves on the steering, organizing, and program committees of several international conferences. He is the steering committee chair/vice-chair for the IEEE International Symposium on Smart Electronic Systems (IEEE-iSES), the IEEE-CS Symposium on VLSI (ISVLSI), and the OITS International Conference on Information Technology (OCIT). He has supervised 3 post-doctoral researchers, 21 Ph.D. dissertations, 29 M.S. theses, and 41 undergraduate projects.

\end{document}